\documentclass[prb,aps,twocolumn,footinbib,floatfix,10pt,longbibliography,superscriptaddress]{revtex4-1}
\usepackage[english]{babel}
\usepackage{breakcites}

\usepackage[utf8]{inputenc}
\usepackage[T1]{fontenc}

\usepackage{graphicx}
\usepackage{braket}
\usepackage{color}
\usepackage{epstopdf}
\usepackage{mathtools}
\usepackage{amsmath}
\usepackage{amssymb}
\usepackage{float}
\graphicspath{{./figs/}}
\usepackage{gensymb}
\usepackage[colorinlistoftodos,prependcaption]{todonotes}
\usepackage{xargs}  
\newcommandx{\unsure}[2][1=]{\todo[linecolor=red,backgroundcolor=red!25,bordercolor=red,#1]{#2}}
\newcommandx{\change}[2][1=]{\todo[linecolor=blue,backgroundcolor=blue!25,bordercolor=blue,#1]{#2}}
\newcommandx{\info}[2][1=]{\todo[linecolor=OliveGreen,backgroundcolor=OliveGreen!25,bordercolor=OliveGreen,#1]{#2}}
\newcommandx{\improvement}[2][1=]{\todo[linecolor=Plum,backgroundcolor=Plum!25,bordercolor=Plum,#1]{#2}}
\newcommandx{\thiswillnotshow}[2][1=]{\todo[disable,#1]{#2}}
\newcommandx{\greencom}[2][1=]
{\todo[inline, color=green!40,#1]{#2}}
\newcommandx{\bluecom}[2][1=]
{\todo[inline, color=blue!40,#1]{#2}}

\usepackage[colorlinks]{hyperref}
\definecolor{winered}{rgb}{0.5,0,0}
\hypersetup
{
colorlinks=true,
linkcolor=winered,
urlcolor={winered},
filecolor={winered},
citecolor={winered},
allcolors={winered}
}

\usepackage{letltxmacro}
\LetLtxMacro{\ORIGselectlanguage}{\selectlanguage}
\makeatletter
\DeclareRobustCommand{\selectlanguage}[1]{%
  \@ifundefined{alias@\string#1}
    {\ORIGselectlanguage{#1}}
    {\begingroup\edef\x{\endgroup
       \noexpand\ORIGselectlanguage{\@nameuse{alias@#1}}}\x}%
}
\newcommand{\definelanguagealias}[2]{%
  \@namedef{alias@#1}{#2}%
}
\makeatother

\definelanguagealias{en}{english}
\definelanguagealias{EN}{english}

\begin{document}
\sloppy
\emergencystretch 3em
\raggedbottom

\newcommand{\sigp}{\sigma^+}
\newcommand{\sigm}{\sigma^-}
\newcommand{\Gspph}{\Gamma^{\sigma^+}_0}
\newcommand{\Gsmph}{\Gamma^{\sigma^-}_0}
\newcommand{\Gcdph}{\Gamma^\mathrm{cd}_0}
\newcommand{\vc}[1]{{\boldsymbol{\mathrm{#1}}}}
\newcommand{\comment}[1]{}
\newcommand{\remove}[1]{}
\newcommand{\quot}[1]{\textquotedblleft{}#1\textquotedblright}
\newcommand{\un}{\mathrm}
\newcommand{\B}{\langle B \rangle}
\newcommand{\blue}[1]{{\color{blue}#1}}
\newcommand{\red}[1]{{\color{red}#1}}
\newcommand{\orange}[1]{{\color{orange}#1}}
\newcommand{\magenta}[1]{{\color{magenta}#1}}
\makeatletter
\newcommand{\vast}{\bBigg@{4}}
\newcommand{\Vast}{\bBigg@{5}}
\newcommand{\G}{\mathbf{G}}
\makeatother
\title{Theory and limits of on-demand single photon sources using plasmonic resonators:\\ a quantized quasinormal mode approach}
%\author{Chris Gustin}
%\email{c.gustin@queensu.ca}
\author{Stephen Hughes}
\affiliation{\hspace{-40pt}Department of Physics, Engineering Physics, and Astronomy, Queen's University, Kingston, Ontario K7L 3N6, Canada\hspace{-40pt}}
\author{Sebastian~Franke}
\affiliation{Technische Universit\"at Berlin, Institut f\"ur Theoretische Physik,
Nichtlineare Optik und Quantenelektronik, Hardenbergstra{\ss}e 36, 10623 Berlin, Germany}
\author{Chris Gustin} 
\affiliation{\hspace{-40pt}Department of Physics, Engineering Physics, and Astronomy, Queen's University, Kingston, Ontario K7L 3N6, Canada\hspace{-40pt}}
\author{Mohsen Kamandar Dezfouli} 
\affiliation{\hspace{-40pt}Department of Physics, Engineering Physics, and Astronomy, Queen's University, Kingston, Ontario K7L 3N6, Canada\hspace{-40pt}}
 \author{Andreas Knorr}
 \affiliation{Technische Universit\"at Berlin, Institut f\"ur Theoretische Physik,
 Nichtlineare Optik und Quantenelektronik, Hardenbergstra{\ss}e 36, 10623 Berlin, Germany}
   \author{Marten Richter}
 \affiliation{Technische Universit\"at Berlin, Institut f\"ur Theoretische Physik,
 Nichtlineare Optik und Quantenelektronik, Hardenbergstra{\ss}e 36, 10623 Berlin, Germany}
\date{\today}

\begin{abstract}
Quantum emitters coupled to plasmonic resonators are known to allow enhanced  broadband 
Purcell factors, and such systems have been recently suggested as possible candidates
for on-demand single photon sources, with fast operation speeds.
However, a true single photon source
has strict requirements of high efficiency (brightness) and quantum indistinguishability of the emitted photons, which can be quantified through  two-photon interference experiments.
To help address this problem, 
we employ and extend a recently developed quantized
quasinormal mode approach,
which rigorously quantizes
arbitrarily lossy open system modes,
to compute the key parameters
that accurately quantify the figures of merit for plasmon-based single
photon sources. We also present a quantized input-output theory
to quantify the radiative and nonradiative quantum efficiencies.
We  exemplify the theory using a nanoplasmonic dimer resonator
made up of two gold nanorods,
which yields large Purcell factors and 
good radiative output beta factors.
Considering an optically pulsed excitation scheme, we explore the key roles of pulse duration  and pure dephasing on the single photon properties, and show that 
ultrashort pulses  (sub-ps)
are generally required for such structures, even for
low temperature operation. We also quantify the role of the nonradiative beta factor both for single photon and two-photon emission processes. 
Our general approach can be applied to a wide variety of plasmon systems, including metal-dielectrics,  and cavity-waveguide systems, without recourse to phenomenological
quantization schemes. 
%The theory is also applicable to various dielectric systems, especially ones where the cavity mode quality factor is low. 
\end{abstract}
\maketitle

% TO DO
% \greencom{{\bf To do:} \\ fix fig 5 label (pulse to p) and check FOM is clear with impact of beta factor in definitions on graphs and discussions}

\section{Introduction}\label{sec0}

It is now well established that a
 single two-level system (TLS)
 coupled to a resonant-cavity can lead  to the creation of single photons ``on demand''~\cite{senellart17,kuhn10,muller17,broome13}, which can be exploited for applications in quantum photonics as well as for fundamental quantum optical studies. The essential requirements for a practical on-demand single photon source (SPS), include
the ``brightness'' or efficiency (probability of emitting a single photon per single excitation pulse) as well as the
``indistinguishability'', which is critical for
two-photon interference and quantum protocols.
Recently, there have been impressive improvements made with quantum dots (QDs) coupled to semiconductor cavity systems~\cite{buckley12,ding16,somaschi16,he17,liu17},
where the QD  mimics an artificial atom or TLS. The coupled cavity-QD system allows an increase in the radiative decay rate through the Purcell effect, which helps to mitigate the detrimental effects of pure dephasing
and excitation-induced dephasing processes, e.g., caused by electron-phonon
interactions~\cite{lodahl15,nazir16,roy11,forstner03,reiter2014}.

Nanoplasmonics offers another class of cavity systems that 
can enhance the light-matter interactions for SPS applications, and there have been various works published on how to increase the 
Purcell factor (enhanced spontaneous emission factor)
as well as on how to increase the output (radiative) beta factor. For example,
Belacel {\em et al.}~\cite{belacel_controlling_2013}
experimentally showed how to increase the Purcell factor
above 80 using metal nanopatch entennas.
Siampour {\em et al.}~\cite{siampour_nanofabrication_2017}
fabricated a plasmonic circuit, using
NV centers coupled to surface plasmon waveguides,
demonstrating a 5-fold increase in the 
spontaneous emission rate, and more than
50\% coupling to an output waveguide mode.
Bulu {\em et al.}~\cite{bulu_plasmonic_2011}
described how NV centers coupled to plasmonic
resonators can yield Purcell factors
of 50 and collection efficiencies of more than 40\%.
Huang {\em et al.}~\cite{hoang_ultrafast_2016-1}
demonstrated room-temperature single photon emission from QDs coupled to plasmonic nanocavities, using silver nanocubes,
and measured impressive Purcell factors in excess of 500; they also
showed how the nanocavity acts as a highly efficient optical antenna directing the emission into a single
lobe normal to the surface.
Lyamkina {\em et al.}~\cite{lyamkina_monolithically_2016}
showed how one can
deterministically integrate semiconductor quantum emitters with plasmonic nano-devices, useful for 
chip-scale integration and true nanoscale quantum photonic technologies; they    demonstrated strong enhanced light-matter coupling (enhanced spontaneous emission) of single near-surface ($<$10 nm) InAs QDs monolithically integrated into electromagnetic hot-spots of sub-wavelength sized metal nanoantennas. 
Most of these papers agree that two of the most important metric for efficient SPS from metals are the Purcell factor and output beta factor or quantum efficiency. In plasmonics, these
two properties are not mutually exclusive, even at the level of a single mode.

In terms of measurements of the single photon purity, there
have also been various experiments with metal resonators.
 Livneh {\em et al.}~\cite{livneh_highly_2016} 
 demonstrated a room temperature SPS features from
 a  bulls-eye shaped hybrid metal-dielectric nanoantenna,
 and suggested that such a device 
 ``paves a promising route for a high purity, high efficiency, on-chip single photon source operating at room temperature.'' In this experiment
 they also measured antibunching 
 using off-resonant pulses excitation, with a pulse duration of 55\,ps.
 Liu {\em et al}.~\cite{liu_bright_2013} showed
bright SPS based on an InAs quantum dot in a silver-embedded nanocone structure;
they demonstrated
 photon emission rates of $\sim$200,000 photons per second and single-photon emission with autocorrelation measurements, using 5 ps excitation pulses;
further, suppression of multi-photon emission with $g^{(2)}(0){<}0.1$ was achieved by quasi-resonant excitation at low excitation power. 

However, to the best of our knowledge there has been no reported measurements of the two-photon interference for metal based SPSs, which is required to quantify the indistinguishability of the emitted photons. The common metric of showing $g^{(2)}(0) \ll 0.5$
demonstrates a good quantum emitter (single TLS), but this information alone is not enough to quantify a good SPS, even if $g^{(2)}(0) = 0$. On the other hand,
the tremendous progress over the past few years on QD SPS, has been largely based on suppressing charge noise (pure dephasing of the zero phonon line),
resonant driving (which avoids unwanted background excitations), and using the optimal pulse duration and Purcell factors, so the design requirements for these systems are now clearer.
Most recent experiments with QDs in semiconductor cavities and plasmonics use pulses on the picosecond timescale, typically 2-12 ps~\cite{somaschi_near-optimal_2016}. However, it has been pointed out that the often neglected role of the excitation pulse can cause additional dephasing and result in 
multi-photon emission per excitation pulse~\cite{fischer17,gustin_pulsed_2018}.
Thus, generally one should work in the following pulse excitation regime:
$\tau_{\rm p} {\ll} 1/ \gamma^{\rm P}$,
where $\tau_{\rm p}$ is the pulse duration, and $\gamma^{\rm P} {=} {F}_{\rm P}  \gamma$ with $\gamma$ as the (non-cavity) spontaneous emission rate and 
${F}_{\rm P}$ as the cavity-enhanced Purcell factor; 
however, this requirement can be relaxed if $\tau_{\rm p} \kappa {\ll} 1$, where $\kappa$ is the cavity photon decay rate.
This places an upper limit on the pulse duration. On the other hand, increasing pure dephasing of the quantum emitter (QD) is also known to spoil
the SPS properties, so it is not clear how one could ever get room temperature based systems, yet this is often cited as one of the advantages of plasmon-based quantum light sources, as well as faster operation speeds and smaller spatial footprints.

In metal-based SPS applications, the common trend has been to increase the  Purcell factor
and the output beta factor (to partly combat intrinsic material losses). Unfortunately, this does not necessarily say anything
about the indistinguishability of the emitted photons, so it is not clear if one can  realize the possible advantages, namely high temperate operation and ultrafast speed. Also, a large Purcell factor can sometimes lead to greater multi-photon pair emission, even in a bad cavity limit. A further complication is
the often {\em ad hoc} quantization of the underlying plasmon modes. The most common treatment
uses a driven Jaynes-Cumming (JC) model~\cite{PhysRevA.82.043845,PhysRevLett.105.263601,rousseaux_comparative_2018}, 
which has been used to compare different designs that
are assumed to behave like a single mode
JC model, then computing quantities like the degree of antibunching from
a continuous wave (CW) driven system. 
Thus there is now a pressing need for a more fundamental theory for metallic resonators for SPS applications, 
and there is a rather urgent need to clarify and quantify the most useful figures of merit for plasmonic SPSs.

The use of mode theories for plasmonics
has been hotly debated over the years~\cite{koenderink_use_2010}.
However,  recently the use of quasinormal modes (QNMs)~\cite{lee_dyadic_1999,kristensen_generalized_2012,sauvan_theory_2013,kristensen_modes_2014,ge_quantum_2015,noauthor_light_nodate} have proven very successful to calculate the generalized effective mode volumes, Purcell factors, and
the photon Green's function for use in system-reservoir theory of quantum optics.
Such an insightful modal approach yields quantitatively good agreement with full dipole simulations
in the bad cavity limit, but the quantization of such modes is challenging.
 Very recently, a formal derivation of a QNM quantum master equation was introduced~\cite{franke_quantization_2018}, which allows one to rigorously study multi-photon problems
for metallic resonators with completely derived mesoscopic coupling elements, even with very large material losses.

In this paper, we exploit a quantized QNM master equation
to study the key figures of merit for metal resonator
SPSs. We also extend the theory to obtain the required output field operators. We remark that several deviations to a phenomenological approach appear when using the quantized QNM theory. For instance, the inherent dissipative nature of the QNMs requires a renormalization of the field operators, and hence the field-emitter coupling and the dissipation are directly connected to each other and can not be chosen independently. Furthermore, in the case of more then one QNM, dissipation induced coupling terms between different QNMs appear in the Lindblad master equation.
Using a first principles example of a gold metal dimer, we compute the important SPS figures of merit,
and show how the common metrics of Purcell factor and output beta factor show up in the efficiency and indistinguishably. Our approach can be applied to a wide range of plasmonic systems, including systems with coupled modes, and even coupled waveguides~\cite{kamandar_dezfouli_regularized_2018} and hybrid metal-dielectric modes~\cite{kamandar_dezfouli_quantum_2017}.
In Sec.~\ref{sec1}, we first describe the main classical theory and
quantum theory for QNMs, specialized to one QNM. 
We present the QNM master equation and also show how one can obtain the output operators, and spatially averaged correlation functions, and discuss the subtle role of radiative and non-radiative emission. Next,
in Sec.~\ref{sec2},
we present the key figures of merit for SPSs, and clarify the differences for metal based systems from dielectric resonators.
In Sec.~\ref{sec3}, we present the main calculation results for an example gold resonator dimer, whose large Purcell factor and good beta factor is already known classically~\cite{ge_design_2014};
we show the classical QNM properties, and obtain the parameters for the second quantization theory. 
We then examine the SPS figures of merit, with and without pure dephasing and explicitly show how to optimize the pulse duration. In contrast to many claims in the literature, 
we show that the indistinguishabilities are typically not practical for pure dephasing rates in excess of 5 meV (since they are basically classical), unless one works with extremely ultrashort (fs)  pulses and Purcell factors close to $10^5$. For low temperature operation, however, broadband Purcell factors in excess of 1000 can be exploited, but only by using 
sub-ps pulses (e.g., 0.3\,ps if the Purcell factor is around 1500). 
We present a summary in Sec.~\ref{sec5}.

%In Sec.~\ref{sec4}, we present general discussions, and then conclude
%in Sec.~\ref{sec5}.

\section{Quasinormal theory and  quantization}\label{sec1}

\subsection{Classical quasinormal mode theory, Green's functions, Purcell factors, and beta factors}

The open cavity  QNMs, $\tilde{\mathbf{f}}_{{\mu}}\left(\mathbf{r}\right)$,
for any arbitrary resonator system,
 are solutions to the Helmholtz equation,
\begin{equation}
\boldsymbol{\nabla}\times\boldsymbol{\nabla}\times\tilde{\mathbf{f}}_{{\mu}}\left(\mathbf{r}\right)-\left(\dfrac{\tilde{\omega}_{{\mu}}}{c}\right)^{2}\epsilon\left(\mathbf{r},\tilde{\omega}_\mu\right)\,\tilde{\mathbf{f}}_{{\mu}}\left(\mathbf{r}\right)=0,
\end{equation}
 subject to open boundary conditions,
i.e.,  the Silver-M\"uller radiation condition \cite{Kristensen2015}; here $\epsilon\left(\mathbf{r},\omega\right)$  is the complex dielectric function of the system and $\tilde{\omega}_{{\mu}}{=} \omega_{{\mu}}{-}i\kappa_{{\mu}}/2$
 is the complex resonance frequency that can also be used to quantify the
modal quality factor as $Q_{\mu}=\omega_{\mu}/\kappa_{\mu}$, where 
$\kappa_{\mu}$ is the full-width half-maximum decay rate. The QNMs, once normalized, can be used to construct the transverse Green function through~\cite{leung_completeness_1994,ge_quasinormal_2014}
\begin{equation}
\mathbf{G}\left(\mathbf{r},\mathbf{r}_{0};\omega\right)= \sum_{\mu} A_{\mu}\left(\omega\right)\,\tilde{\mathbf{f}}_{\mu}\left(\mathbf{r}\right)\tilde{\mathbf{f}}_{\mu}\left(\mathbf{r}_{0}\right),\label{eq:GFwithSUM}
\end{equation}
for locations nearby or within the  scattering geometry, where the QNMs can form a complete basis~\cite{leung_time-independent_1994,leung_completeness_1996}, and  we have defined the spectral function
\begin{equation}
A_{\mu}(\omega)=\frac{\omega}
{2\left(\tilde{\omega}_{\mu}-\omega\right)}.
\end{equation}
%\begin{equation}
%A_{\mu}(\omega)=\frac{\omega^{2}}
%{2\,\tilde{\omega}_{\mu}\left(\tilde{\omega}_{\mu%}-\omega\right)}.
%\end{equation}
%
The total photon Green's function, ${\bf G}(\mathbf{r},\mathbf{r}';\omega)$,
%is a solution of
%where ${\bf G}(\mathbf{r},\mathbf{r}';\omega)$ is the photonic Green's function and 
fulfills 
\begin{equation}
\nabla\times\nabla\times{\bf G}(\mathbf{r},\mathbf{r}';\omega)-\frac{\omega^2}{c^2}\epsilon(\mathbf{r},\omega){\bf G}(\mathbf{r},\mathbf{r}';\omega)=\frac{\omega^2}{c^2}\mathbf{1}\delta(\mathbf{r}-\mathbf{r}'),\label{eq:GreenHelmholtz}
\end{equation}
with suitable radiation conditions.

We next consider a single,
QNM, $\mu{=}c$, which is the most practical case
for SPS designs.
 The single QNM Green function can thus be written as
\begin{equation}
\mathbf{G}_{\rm c}\left(\mathbf{r},\mathbf{r}_{0};\omega\right) \approx A_{\rm c}(\omega)\,\tilde{\mathbf{f}}_{\rm c}\left(\mathbf{r}\right)\tilde{\mathbf{f}}_{\rm c}\left(\mathbf{r}_{0}\right),
\label{eq:GF_QNM}
\end{equation}
where again this  holds only nearby the cavity region (typically at distances where a quantum emitter still feels a reasonable Purcell factor enhancement).

For any finite radiation leakage, the QNM spatially diverges at locations far away
from the resonator~\cite{kristensen_generalized_2012},  and this divergence depends upon the value
of $Q_{\rm c}$.
The divergent behavior of the
total field is of course unphysical, as we know there should be no enhanced emission in the far field, and the total field must be convergent. One solution to this problem is to use 
 a Dyson equation formalism to reconstruct the full Green function at locations away (outside) from the cavity region \cite{ge_quasinormal_2014}, so that  one can obtain a ``regularized'' mode form,
\begin{align}
\tilde{\mathbf{F}}_{\rm c}(\mathbf{R}) = \int_{\rm cavity} \mathbf{G}_{\rm hom}(\mathbf{R},\mathbf{r^{\prime}};\omega)\,\Delta \epsilon(\mathbf{r^{\prime}},\omega)\cdot\tilde{\mathbf{f}}_{\rm c}(\mathbf{r^{\prime}})\,{\rm d} \mathbf{r}',\label{eq:bigF}
\end{align}
for any position ${\bf R}$ outside the resonator, in real frequency space. Here, $\mathbf{G}_{\rm hom}$ is the Green function for the background medium  and $\Delta\epsilon({\bf r},\omega)=\epsilon({\bf r},\omega)-\epsilon_{\rm B}(\bf r)$ is the total dielectric constant minus the background term $\epsilon_{\rm B}(\bf r)$. 
 The regularized mode, $\tilde{\mathbf{F}}_{\rm c}(\mathbf{r})$, can be used in a similar Green function expansion as in Eq.\,\eqref{eq:GF_QNM} to obtain physically meaningful quantities far outside the resonator, where it is known that a single QNM approach will breakdown.
This regularized mode can then be used at all positions outside the  resonator, and has previously been shown to be highly accurate when compared to full dipole calculations \cite{ge_quasinormal_2014}. Indeed,
the enhanced emission can be shown to 
go to one in the far field limit, if using
$\tilde {\bf F}_{\rm c}$.
Another approach can obtain these 
regularized QNMs directly from dipole simulations in real frequency space~\cite{kamandar_dezfouli_regularized_2018} (implemented for for single mode resonators).

Calculation and normalization of QNMs in photonics
has emerged as an important topic in cavity physics, and various methods
have now been demonstrated using both time-domain techniques, such as FDTD, and frequency-domain methods (such as finite-element solvers)
~\cite{kristensen_modes_2014,ge_quasinormal_2014,sauvan_theory_2013}. Using a frequency domain approach, an efficient dipole normalization technique was developed by Bai {\it et al.}~\cite{bai_efficient_2013-1}, implemented using COMSOL \cite{comsol}, where the self-consistent
response to a dipole excitation is used to obtain an {\it integration-free}
normalization for the QNM. 
A similar approach has also been introduced
in FDTD~\cite{kamandar_dezfouli_regularized_2018}, compatible with Lumerical FDTD~\cite{lumerical}. It is also useful to note that QNM
solvers can now model various complex geometries, such as cavities and metals coupled to output waveguides~\cite{malhotra_quasinormal_2016,kamandar_dezfouli_regularized_2018}.

After the QNM Green function is obtained, it is easy to compute the enhanced emission factor of a dipole emitter (and at any position).
 Considering a  dipole
emitter polarized along $\mathbf{n}_d$, placed at position $\mathbf{r}_0$,
the generalized Purcell factor is \cite{Anger2006,kristensen_modes_2014} 
% \begin{equation}
% F_{{\rm P}}({\bf r}_0;\omega)=\frac{6\pi c^{3}}{\omega^{3}n_{\rm B}}\,\mathbf{n}_d\cdot{\rm Im}\{\mathbf{G}_{\rm c}\left(\mathbf{r}_0,\mathbf{r}_0;\omega\right)\}\cdot\mathbf{n}_d,
% \end{equation}
\begin{align}
F_{{\rm P}}({\bf r}_0;\omega) &=
\frac{\mathbf{n}_d\cdot{\rm Im}\{\mathbf{G}_{\rm c}\left(\mathbf{r}_0,\mathbf{r}_0;\omega\right)\}\cdot\mathbf{n}_d}{\mathbf{n}_d\cdot{\rm Im}\{\mathbf{G}_{\rm hom}\left(\mathbf{r}_0,\mathbf{r}_0;\omega\right)\}\cdot\mathbf{n}_d} \nonumber \\
&=\frac{6\pi c^{3}}{\omega^{3}n_{\rm B}}\,\mathbf{n}_d\cdot{\rm Im}\{\mathbf{G}_{\rm c}\left(\mathbf{r}_0,\mathbf{r}_0;\omega\right)\}\cdot\mathbf{n}_d,
\end{align}
where $n_{\rm B}$ is the background refractive index
and ${\rm Im}[{\bf G}_{\rm hom}
({\bf r},{\bf r})]_{ii}{=}n_{\rm B}\omega^3/6\pi c^3$.
Note that for dipoles outside the resonator,
the total rate is $F_{{\rm P}}({\bf r};\omega)+1$, where the factor of 1 comes from using scattering theory for dipoles outside the metal~\cite{ge_quasinormal_2014} (essentially the contribution from the homogeneous radiation modes). 
The generalized effective mode volumes at the dipole location \cite{kristensen_generalized_2012}, is
\begin{equation}
{\rm V_{eff}} = \frac{1}{{\rm Re}\{\epsilon\left(\mathbf{r}_0\right)\tilde{\mathbf{f}}_{\rm c}^2\left(\mathbf{r}_0\right)\}},
\end{equation}
which can then be used in the  Purcell factor
definition~\cite{kristensen_modes_2014}:
\begin{equation}
F_{\rm P}
= \frac{3}{4\pi}\left(\frac{\lambda}{n_{\rm B}}\right)^3
\frac{Q}{V_{\rm eff}} \eta({\bf n}_d,{\bf r}_0,\omega).
\end{equation}
Since the usual Purcell formula  is based on the assumption that the
emitter is at the field maximum (both spectrally and spatially)
and that the dipole moment orientation is parallel to the field
at this point, the above is a simple generalization to account
for any deviation with the factor 
$\eta({\bf n}_d,{\bf r}_0,\omega)$~\cite{kristensen_modes_2014}.
Typically such a factor is quoted on-resonance, where
$\omega{=}\omega_{\rm c}$, unless shown as a function of frequency.

% where $\eta$ accounts for a change in position and polarization 
% with respect to the local field maximum  and 
% orientation of the QNM polarization; 
% for maximum coupling, 
% $\eta=1$.
%However, we prefer to keep the more general
%expression above.

It is important to stress that both radiative and nonradiative contributions come from the same QNM, so both of these contributions scale as $Q/V_{\rm eff}$ with $Q{=}Q_{\rm c}$. It is sometimes common to assume that  the plasmon mode emission is through $\gamma_{\rm rad}$,
and the ``other'' channels are through non-radiative loss, but there is usually no need to add any extra modes at all in QNM
theory (unless the dipole is very near the metal walls, where evanescent modes play a more significant role). 
Indeed, a single QNM can be rigorously
valid even for few nm gap antennas~\cite{KamandarDezfouli2017}, and nonradiative coupling is part of the mode
eigenvalue solution.
Thus, the QNM is intimately related to both radiative and non-radiative decay,
and  the total single QNM spontaneous emission rate
depends on the dipole strength, $d$.

Below,  we use single QNM theory, since it is the most practical case for SPS sources; thus we will drop the
`$\mu{=}{\rm c}$' labelling,  as the single mode is assumed to be implicit, e.g.,
%$\beta_{\rm rad} =\beta_{\rm rad}^{\rm c},
$\tilde{\mathbf{f}}{=}
\tilde{\mathbf{f}}_{\rm c},$ etc. We are also assuming dipole positions that are dominated by the single QNM properties, which we also rigorously justify and confirm below, for the chosen metal dimer
structure.
The total TLS decay rate is defined from:
% so that \red{just a note: if we change $\gamma$
% to $\Gamma$ here, we also need to do in the ME and all the graphs that list
% $\gamma'$; but this is easy to change also, so I am fine with either notation
% as long as we define them properly}\blue{SF: I think best way is to change $\gamma^{\rm c}$ to $\gamma^{\rm P}=F_{\rm p}\gamma$, as this is also used at the beginning of Section IV,B. I would also change $\beta_{\rm c}$ to $\beta$. In the special case of one QNM, the QNM decay rate $\kappa$ is related to $beta$, but this is only true for a single QNM. The relations below are quiet general for cavity-enhanced spontaneous emission.}
\begin{equation}
\gamma^{\rm tot} = \frac{2 d^2 \mathbf{n}_d\cdot{\rm Im}\{\mathbf{G}\left(\mathbf{r}_0,\mathbf{r}_0;\omega\right)\}\cdot\mathbf{n}_d}{\varepsilon_0\hbar},
\end{equation}
which, in the limit of a single QNM response (${\bf G}\rightarrow {\bf G}_{\rm c}$), can be written
as a cavity-enhanced (or modified) emission rate:
\begin{equation}
\gamma^{\rm tot} \rightarrow  \gamma^{\rm P} = F_{\rm P} \gamma,
\end{equation}
as discussed in the introduction,
where $\gamma$ is the homogeneous background emitter decay rate
and $F_{\rm P}$ is the single QNM Purcell factor.
For metal antenna systems, it is important to stress that this cavity-enhanced decay rate
 has both radiative and non-radiative parts: 
\begin{equation}
\gamma^{\rm P} = \gamma_{\rm rad}^{\rm P}+\gamma_{\rm nrad}^{\rm P}.
\end{equation}
 Consequently, the other important quantity that can be obtained from the classical QNM theory,
is the radiative beta factor, defined as
\begin{equation}
\beta_{\rm rad} = \frac{\gamma_{\rm rad}^{\rm P}}{\gamma_{\rm rad}^{\rm P}
+\gamma_{\rm nrad}^{\rm P}}.\label{eq:ClassBetarad}
\end{equation}
This gives the probability that the cavity emitted photon will 
be radiatively emitted out of the antenna system.
Note $\gamma_{\rm nrad}^{\rm P}$
can also be computed from
the QNM mode~\cite{Anger2006,sauvan_theory_2013,KamandarDezfouli2017}, or directly in any Maxwell solver by integrating the power flow (if in the regime of
a single QNM response). The non-radiative beta factor is defined similarly:
\begin{equation}
\beta_{\rm nrad} = \frac{\gamma_{\rm nrad}^{\rm P}}{\gamma_{\rm rad}^{\rm P}
+\gamma_{\rm nrad}^{\rm P}},\label{eq:ClassBetanrad}
\end{equation}
where $\beta_{\rm nrad} {+} \beta_{\rm nrad}{=}1$. We also remark that  this is the ideal case that  neglects non-radiative losses in the TLS, though such processes could also be included in the definitions above.

\subsection{Quantized quasinormal mode theory}

%and input-output formalism
%for the output field operators}

Starting from a Green-function quantization scheme for an inhomogeneous and dissipative medium, the Hamiltonian of the medium-assisted electric field coupled to a TLS, 
reads~\cite{Dung,Suttorp}
\begin{align}
H_{\text{total}}=& \hbar \omega_a\sigma^+\sigma^- + \hbar\int {\rm d}\mathbf{r}\!\int_0^{\infty}\!{\rm d}\omega~\omega~\hat{\mathbf{b}}^{\dagger}
(\mathbf{r},\omega)\cdot\hat{\mathbf{b}}(\mathbf{r},\omega)\nonumber \\
&-\left[\sigma^+ \int_0^{\infty}{\rm d}\omega~\mathbf{d}_a\cdot\hat{\mathbf{E}}(\mathbf{r}_0,\omega)+\text{H.a.}\right],
\label{eq:totham}
\end{align}
where $\omega_a$ and $\mathbf{d}_a{=}d {\bf n}_d$ are the TLS angular resonance frequency and dipole moment, respectively, $\sigma^\pm$ denote TLS raising and lowering operators, and we use a dipole-field interaction in the rotating wave approximation (RWA);
the annihilation and creation operators $\hat{\mathbf{b}}(\mathbf{r},\omega)$ and $\hat{\mathbf{b}}^{\dagger}(\mathbf{r},\omega)$ act on joint excitations of the surrounding lossy media (plasmons) 
and electromagnetic degrees of freedom (photons) and
satisfy  canonical commutation relations~\cite{Dung}. We note, that Eq.~\eqref{eq:totham} is given in Schr\"odinger picture, and $\omega$ is a mode index, rather then a Fourier variable of time. Indeed, in a Heisenberg picture, one has $\hat{\mathbf{E}}(\mathbf{r}_0,\omega,t)$.

The electric field operator $\hat{\mathbf{E}}(\mathbf{r},\omega)$ is the solution to the Helmholtz equation,
\begin{equation}
\nabla {\times}  \nabla {\times}{\hat{\mathbf{E}}}(\mathbf{r},\omega)
-\frac{\omega^2}{c^2}\epsilon(\mathbf{r},\omega){\hat{\mathbf{E}}}(\mathbf{r},\omega)=
i\omega\mu_0 {\hat{\mathbf{j}}}_{\text{noise}}(\mathbf{r},\omega),\label{eq:Helmholtz}
\end{equation}
where $\hat{\mathbf{j}}_{\text{noise}}(\mathbf{r},\omega){=}\omega
\sqrt{(\hbar\epsilon_0/\pi)\epsilon_{I}(\mathbf{r},\omega)}\hat{{\mathbf{b}}}(\mathbf{r},\omega)$ is the noise operator associated with absorption and radiative loss of the system. The source-field representation of Eq.~\eqref{eq:Helmholtz} is given via
\begin{equation}
{\hat{\mathbf{E}}}(\mathbf{r},\omega)=\frac{i}{\omega\epsilon_0}\int d \mathbf{r}' {\bf G}(\mathbf{r},\mathbf{r}';\omega)\cdot{\hat{\mathbf{j}}}_{\text{noise}}(\mathbf{r}',\omega)\label{eq:Sourcefield},
\end{equation}
using the same classical Green's function 
as defined through Eq.~\eqref{eq:GreenHelmholtz}.

As shown in Ref.~\onlinecite{franke_quantization_2018}, using the QNM Green's function in Eq.~\eqref{eq:GFwithSUM} in combination with the source field expression in Eq.~\eqref{eq:Sourcefield}, one can expand the medium-assisted electric field in terms of QNM operators. Below we focus on the special case of this expansion for one QNM (since it is the most practical case for SPS emitters), as also discussed in the classical QNM theory above. The total electric field operator, ${\hat{\mathbf{E}}}(\mathbf{r}_{\rm s}){=}\int_0^{\infty}{\rm d}\omega{\hat{\mathbf{E}}}(\mathbf{r}_{\rm s},\omega){+}{\rm H.a.}$, for positions $\mathbf{r}_{\rm s}$ (system region), reads as 
\begin{equation}
\hat{\mathbf{E}}(\mathbf{r}_{\rm s})=i\sqrt{\frac{\hbar\omega_{\rm c} }{2\epsilon_0  }}\, \sqrt{S}\tilde{\mathbf{f}}_{}(\mathbf{r}_{\rm s}) a + \text{H.a.},\label{eq:Esymm}
\end{equation}
 in a symmetrized basis, with
 \begin{equation}
    a=\sqrt{\frac{2}{\pi\omega_{\rm c}S}}\int_0^{\infty}{\rm d}\omega A_{\rm c}(\omega)\int{\rm d}\mathbf{r}\sqrt{\epsilon_I(\mathbf{r},\omega)}\tilde{\mathbf{f}}_{}(\mathbf{r})\cdot\hat{\mathbf{b}}(\mathbf{r},\omega)\label{eq:QNMoperator},
\end{equation}
where  $a$ and $a^\dagger$
 are suitable annihilation and creation operators to obtain plasmon/photon Fock states for the symmetrized QNM $ \sqrt{S}\tilde{\mathbf{f}}(\mathbf{r})$. The $S$
 ``photon normalization factor'' is obtained from
\begin{equation}
S=\frac{2}{\pi\omega_{\rm c}}\int_0^{\infty}\!{\rm d}\omega |A_{\rm c}(\omega)|^2\left[S^{\rm nrad}(\omega){+}S^{\rm rad}(\omega)\right]\label{eq:S-definition},
\end{equation}
where 
\begin{equation}
S^{\rm nrad}(\omega)=\int_{V} {\rm d}\mathbf{r}\,\epsilon_I(\mathbf{r},\omega)\,
|\tilde {\mathbf{f}}({\bf r})|^2,
\end{equation}
accounts for absorption due to the metallic resonator material, and   
\begin{equation}
S^{\rm rad}(\omega)= \frac{1}{\epsilon_0\omega}\int_{S_{V}} {\rm d}A_{\mathbf{s}}\mathbf{n}_{\mathbf{s}}\cdot {\rm Re}(\tilde{\mathbf{F}}(\mathbf{s},\omega)\times\tilde{\mathbf{H}}^*(\mathbf{s},\omega))
\end{equation}
describes radiation leaving the system through the surface $S_V$ with the normal vector $\hat{\mathbf{n}}_{\mathbf{s}}$ pointing into the resonator volume $V$, and $\tilde{\mathbf{H}}(\mathbf{s},\omega){=}i/(\mu_0\omega)\nabla
\times\tilde{\mathbf{F}}(\mathbf{s},\omega)$ is the QNM magnetic 
field. We note here, that $S^{\rm rad}(\omega)$ looks similar to the classical (radiative) power flow (cf. Ref.~\onlinecite{liberal2018control}) as an integral over the (time-averaged) Poynting vector, which in terms of the classical fields yields 
$\mathbf{S}_{\rm Poynting}{=}0.5{\rm Re}(\mathbf{E}(\omega)\times\mathbf{H}^*(\omega))$.
The $S^{\rm rad}(\omega)$ thus has a clear interpretation: it is the normalized QNM power flow outside the antenna and it appears naturally in our formalism. It is important to also stress that we obtain this factor even in the limit of a lossless resonator structure, where in general a QNM quantization is still required, especially for
low $Q$ cavities; in this case, the $S$ expression is exactly the same, though 
$S^{\rm nrad}(\omega){=}0$. However, in general this latter contribution will, in general, not be zero at certain frequencies, as is required through causality and the Kramers Kronig relations.

The radiative and non-radiative contributions are connected to the $\beta$ factors, defined also from classical calculations in Eq.~\eqref{eq:ClassBetarad} and Eq.~\eqref{eq:ClassBetanrad};
in our quantum theory, they are defined as $\beta_{\rm rad}{=}S^{\rm rad} /S$ and $\beta_{\rm nrad}{=}S^{\rm nrad}/S$. Since the $\omega$-dependent terms in Eq.~\eqref{eq:S-definition} are dominated by a Lorentzian term around the QNM frequency $\omega_c$, we use a resonance approximation and shift the lower frequency integral limit to $-\infty$, which yields
\begin{gather}
    S^{\rm nrad}\approx Q \int_{V} {\rm d}\mathbf{r}\,\epsilon_I(\mathbf{r},\omega_{\rm c})\,
|\tilde {\mathbf{f}}({\bf r})|^2,\label{eq:Snradapprox}\\
S^{\rm rad}\approx \frac{n_{\rm B}c}{\kappa}\int_{S_{\infty}} {\rm d}A_{\mathbf{s}}|\tilde{\mathbf{F}} (\mathbf{s},\omega_{\rm c})|^2\label{eq:Sradapprox},
\end{gather}
where we evaluated the surface integral in the far field ($S_\infty$) and used the Silver-M\"uller radiation condition on $\tilde{\mathbf{F}} (\mathbf{s},\omega_{\rm c})$ for outgoing fields, i.e., 
%\begin{equation}
    $\hat{\mathbf{n}}_{\mathbf{s}}  {\times} \tilde{\mathbf{H}} (\mathbf{s},\omega){\rightarrow}\, {-}n_{\rm B}c\epsilon_0\tilde{\mathbf{F}} (\mathbf{s},\omega)$. We remark, that $S$ is independent of the choice of $V$ and its surrounding surface $S_\infty$, as long as $S_\infty$ is far enough away from the absorptive region.
These $S$ factors are unitless quantities, and for single QNMs, we usually find~\cite{franke_quantization_2018} that $S{\approx}1$ (see also calculations below). Thus they have a close connection to the quantum efficiencies of radiative and non-radiative emission,
or the SPS beta factors.
%\end{equation}
%\blue{write this out for clarity to the reader}

%For a simpler notation, from here, we set $\hat{\tilde{\alpha}}^{\rm s}\equiv a$. 

% In the following, we use a single QNM approximation \red{if using the multi-mode version above} to the general expansion in Eq.~\eqref{eq:Esymm}, such that $\mu\!=\!\nu\!=\!c$, with  $\hat{\mathbf{E}}(\mathbf{r})$=$i\sqrt{\hbar\omega_c/(2\epsilon_0)}
% \tilde{\mathbf{f}}_{c}^{\rm s}(\mathbf{r})  \hat{\tilde{\alpha}}_c^{\rm s} + \text{H.a.,}$ and $\tilde{\mathbf{f}}_{c}^{\rm s}(\mathbf{r}){=}(\mathbf{S}^{\frac{1}{2}})_{c}
% \tilde{\mathbf{f}}_{c}(\mathbf{r})$, where $S_{cc}=S_c$ is a number, which is related to the non-radiative and radiative beta factor via $\beta_{\rm rad}=S_{c}^{\rm rad} /S_{c}$ and $\beta_{\rm nrad}=S_{c}^{\rm nrad}/S_{c}$. For a simpler notation we set $\hat{\tilde{\alpha}}_c^{\rm s}\equiv a$.

\subsection{Input-output theory, quasinormal mode master equation and output electric field operator}

Using the quantized QNM theory above, 
the time evolution of the QNM annihilation operator, $a$, with respect to the coupling of the medium-assisted field and a TLS, is given in the symmetrized QNM basis through~\cite{franke_quantization_2018}
\begin{equation}
\dot{a}=-\frac{i}{\hbar}[a,H_{\rm sys}]-\frac{\kappa}{2}a-\sqrt{\kappa}a_{\rm in},\label{eq:QLE}
\end{equation}
where $H_{\rm sys}{=}H_{\text{em}}{+}H_a{+}H_{I}$  is the system Hamiltonian in the symmetrized QNM basis, $H_{\text{em}}=\hbar\omega_{c}a^{\dagger}a$ and $H_{I}=-i\hbar\sqrt{S}\tilde{g}a\sigma^+ + {\rm H.a.}$  with the QNM-TLS coupling constant $\tilde{g}{=}\sqrt{\omega_{\rm c}/(2\epsilon_0 \hbar)}\, \mathbf{d}_a {\cdot} \tilde{\mathbf{f}} (\mathbf{r}_a)$.
%\red{and $H_{\rm pump}=$ ...to do}
%is the pump term.

Equation~\eqref{eq:QLE} has the form of a quantum Langevin equation (QLE), where apart from the system dynamical terms, two additional contributions appear: a damping term $-(\kappa/2) a$ associated with the QNM radiative and non-radiative decay (recall $\kappa{=}\kappa_{\rm rad}{+}\kappa_{\rm nrad}$) and a noise input operator  $a_{\rm in}{=}{-}i/\sqrt{2\pi\omega_{\rm c}S\kappa}\int_0^{\infty}{\rm d}\omega \sqrt{\omega}\int{\rm d}\mathbf{r}\sqrt{\omega\epsilon_I(\mathbf{r},\omega)}\tilde{\mathbf{f}}(\mathbf{r})\cdot \mathbf{b}(\mathbf{r},\omega)$ (for details, see Ref.~\onlinecite{franke_quantization_2018}), which counteracts the damping and can be regarded as a quantum Langevin force. Indeed, the presence of $a_{\rm in}$ preserves the equal-time commutation relation $[a,a^{\dagger}]{=}1$ at all times. 
In the Markov limit, $[a_{\rm in}(t),a_{\rm in}^{\dagger}(t')]{=}\delta(t{-}t')$, %and we can rewrite Eq.~\eqref{eq:QLE} as 
%\begin{equation}
%\dot{a}\!=\!-\frac{i}{\hbar}[a,H_{\text{sys}}]-\frac{\kappa}{2}a-\sqrt{\kappa}a_{\rm in},\label{eq:QLE3}
%\end{equation}
and Eq.~\eqref{eq:QLE} has the form of the standard QLE~\cite{gardiner_input_1985}, with Markovian input and output, related via 
\begin{equation}
   a_{\rm out}- a_{\rm in}=\sqrt{\kappa}a,\label{eq:inout}
\end{equation}
where $a_{\rm out}$ is the output operator, associated to the quantum noise force in the time-reversed QLE. In the following, we further treat the input operators as white noise in vacuum state, i.e., there is initially zero quanta in the input states, such that $\langle a_{\rm in}(t)a_{\rm in}^{\dagger}(t')\rangle {=} \delta(t{-}t')$ and all other second order correlation functions vanish~\cite{gardiner_input_1985}.

Based on the QLE form shown in Eq.~\eqref{eq:QLE}, %with the additional assumption, that the quantum state associated with the input operators is the vacuum state, 
we can then derive a Lindblad master equation for the QNM-TLS system using  Stratonovich-Ito calculus  to arrive at
\begin{equation}
\dot{\rho}= -\frac{i}{\hbar}\left[\tilde{H}'_{\text{sys}},\rho\right]
+ \frac{\kappa}{2}\mathcal{D}[a]\rho+\frac{\gamma}{2}\mathcal{D}[\sigma^-]+\frac{\gamma '}{2}\mathcal{D}[\sigma^+\sigma^-]\rho\label{masterQNMV2},
\end{equation}
where $\tilde{H}'_{\text{sys}}{=}\tilde{H}_{\text{sys}}{+}H_{\rm pump}$ is the system Hamiltonian in the interaction picture including time-dependent pumping $H_{\rm pump}{=}\hbar\Omega(t)(\sigma^+{+}\sigma^-)$ and rotating in a frame with  $\omega_L$, and $\mathcal{D}[A]{=}2A\rho A^{\dagger}-\{A^{\dagger}A,\rho\}$. In the interaction picture, $\omega_c\rightarrow \omega_c{-}\omega_L$ and $\omega_a \rightarrow \omega_a{-}\omega_L$. Note, that we have phenomenologically included the (free-space or homogeneous background) spontaneous emission ($\gamma$) and pure dephasing ($\gamma'$) process of the TLS in Eq.~\eqref{masterQNMV2}.

To treat photon detection problems, and to simulate measurements
of correlation functions performed outside the dipole and system region, the derivation of the outgoing electric field operator and the connection to the quantum input and output operators is required. In nanoplasmonics, this is often assumed without any form of derivation. Therefore, we start with the positive rotating part of the source-field contribution of the electric field operator, $\hat{\mathbf{E}}^{(+)}(\mathbf{R}){=}\int_0^{\infty}{\rm d}\omega \hat{\mathbf{E}}(\mathbf{R},\omega)$ for positions $\mathbf{R}$ ``outside'' the cavity (system), using again assuming the single QNM Green's function in combination with Eq.~\eqref{eq:Sourcefield}: 
\begin{align}
   \hat{\mathbf{E}}^{(+)}(\mathbf{R},t) &= i\sqrt{\frac{\hbar\omega_{\rm c}}{2\epsilon_0}}\sqrt{S}\int_0^{\infty}{\rm d}\omega\tilde{\mathbf{F}}(\mathbf{R},\omega)\sqrt{\frac{2}{\pi\omega_{\rm c}S}} A_{\rm c}(\omega)\nonumber
   \\
   & \times\int{\rm d}\mathbf{r}\sqrt{\epsilon_I(\mathbf{r},\omega)}\tilde{\mathbf{f}}_{}(\mathbf{r})\cdot\hat{\mathbf{b}}(\mathbf{r},\omega,t).
\end{align}

We see that the $\omega$-dependent terms in $\hat{\mathbf{E}}^{(+)}(\mathbf{R},t)$ are exactly the product of the $\omega$-dependent terms of the system operator $a$ in Eq.~\eqref{eq:QNMoperator} and the regularized QNM $\tilde{\mathbf{F}}(\mathbf{R},\omega)$, which contains all modes $\omega$, propagating out of the resonator. Assuming that the main contribution of the source field contribution is being carried by the QNM frequency $\omega_{\rm c}$, we do a resonance approximation to the scattered QNM field (only), i.e, $\tilde{\mathbf{F}}(\mathbf{R},\omega){\approx}\tilde{\mathbf{F}}(\mathbf{R},\omega_{\rm c})$, to obtain the Markovian  quantum field outside the system,
\begin{equation}
   \hat{\mathbf{E}}^{(+)}(\mathbf{R},t) \approx i\sqrt{\frac{\hbar\omega_{\rm c}}{2\epsilon_0}}\sqrt{S}\tilde{\mathbf{F}}(\mathbf{R},\omega_{\rm c})a(t),
\end{equation}
for the source field contribution ($S$ defined in Eq.~\eqref{eq:S-definition}). This is expected to be an excellent approximation for positions in the far field region away from the system resonator, otherwise one needs the full
Green function response from the emitter to the detector position~\cite{ge_quantum_2015}.

Next,  to connect the source field contribution to the input and output fields, we introduce a normalized QNM function %\blue{``N'' might be confusing with Noise, but we are running out of letters and this is probably better than prime?}
$\tilde{\mathbf{F}}^{\rm N}(\mathbf{s},\omega_c){=}\sqrt{n_{\rm B}c/(S^{\rm rad}\kappa)}\,\tilde{\mathbf{F}}(\mathbf{R},\omega_{\rm c})$ with respect to the far field integration, so that
\begin{align}
  \int_{S_\infty}{\rm d}A_{\mathbf{s}}|\tilde{\mathbf{F}}^{\rm N}(\mathbf{s},\omega_c)|^2 &=\frac{1}{S^{\rm rad}}\frac{n_b c}{\kappa}  \int_{S_\infty}{\rm d}A_{\mathbf{s}}|\tilde{\mathbf{F}}(\mathbf{s},\omega_c)|^2{=}1, \label{eq:normQNMF}
\end{align}
where we used the approximated version of $S^{\rm rad}$ from Eq.~\eqref{eq:Sradapprox}. Subsequently, the electric field operator outside of the structure becomes 
\begin{align}
    \hat{\mathbf{E}}^{(+)}(\mathbf{R},t) &= i\sqrt{\frac{\hbar\omega_{\rm c}\beta_{\rm rad}}{2\epsilon_0n_b c}}S\tilde{\mathbf{F}}^{\rm N}(\mathbf{R},\omega_{\rm c})\sqrt{\kappa }a(t)\nonumber \\
    & =i\sqrt{\frac{\hbar\omega_{\rm c}\beta_{\rm rad}}{2\epsilon_0n_b c}}S\tilde{\mathbf{F}}^{\rm N}(\mathbf{R},\omega_{\rm c})(a_{\rm out}(t)-a_{\rm in}(t)) \nonumber \\
    &\equiv \hat{\mathbf{E}}^{(+)}_{\rm out}(\mathbf{R},t)-\hat{\mathbf{E}}^{(+)}_{\rm in}(\mathbf{R},t), \label{eq:EoutinRad}
\end{align}
where we have used the input-output relation from Eq.~\eqref{eq:inout} to represent $\hat{\mathbf{E}}^{(+)}(\mathbf{R},t)$ as a linear combination of the output electric field operator $\hat{\mathbf{E}}^{(+)}_{\rm out}(\mathbf{R},t)$ and the input electric field operator $\hat{\mathbf{E}}^{(+)}_{\rm in}(\mathbf{R},t)$. Thus,
as expected, the output field is given as a sum of the source field and the input field. We should note here, that the output electric field $\hat{\mathbf{E}}^{(+)}_{\rm out}(\mathbf{R},t)$ represents not only the radiative part of the medium-assisted electric field, but a linear combination of both, non-radiative and radiative, and only the spatial integration over, e.g., a detector volume outside the resonator regime extracts the radiative outcoupling, as will be shown below.

\begin{figure}[hbt]       
\centering\includegraphics[clip,trim=0cm 0cm 0cm 0.cm,width=0.99\columnwidth]{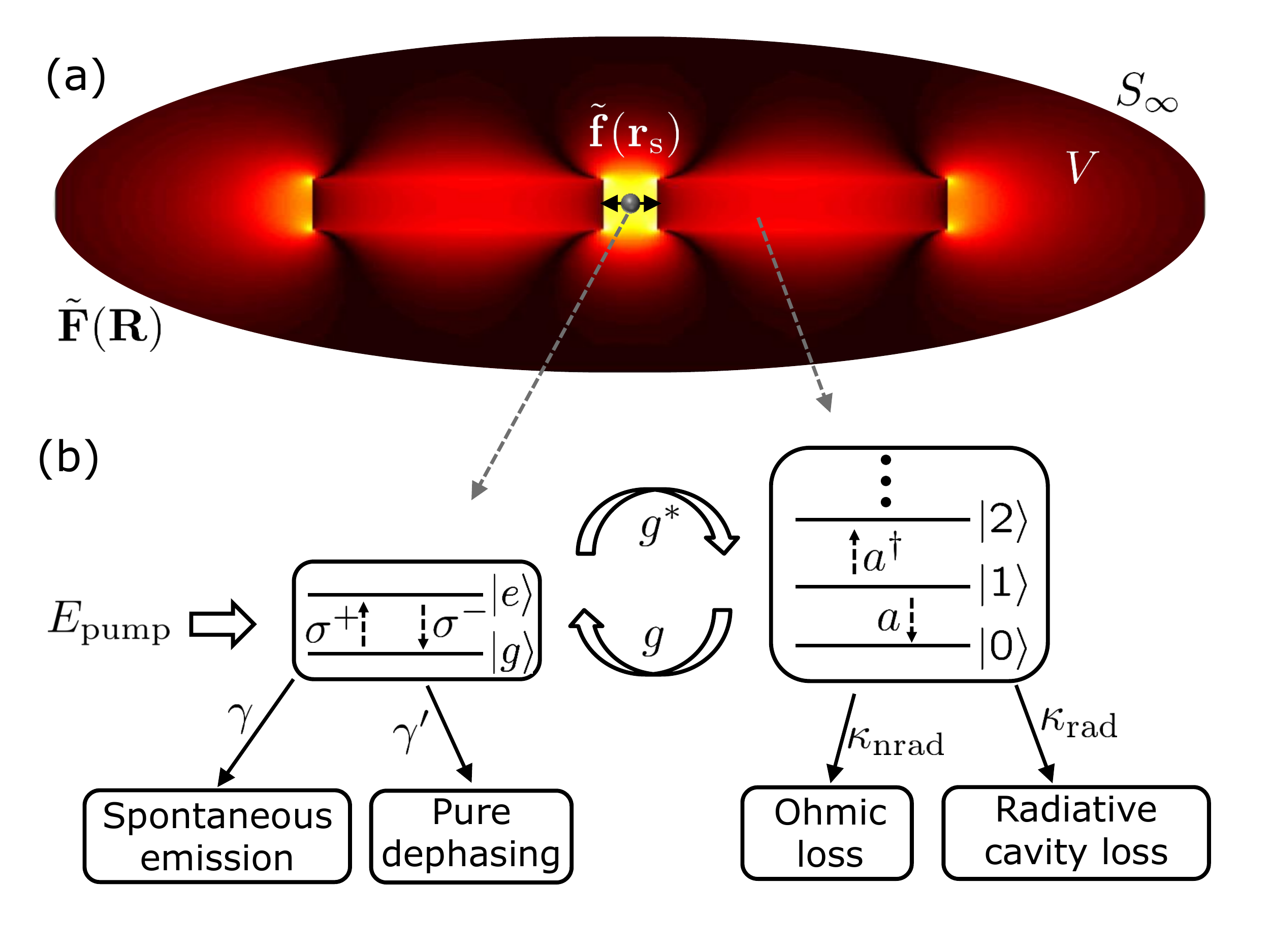}
        \caption{ (a)
        Schematic of a TLS-coupled plasmonic resonator system with virtual cavity volume $V$ and virtual boundary $S_{\infty}$, (b) Visualisation of antenna cavity-QED scenario for a QNM-TLS system with the different decay channels (see text).
 }
        \label{figschematic}
\end{figure}

The same procedure is applied to the electric field for positions $\mathbf{r}_{\rm s}$ inside the metal (i.e., which leads to Ohmic heating); in this case, we start with the electric field operator from Eq.~\eqref{eq:Esymm}, slightly rewritten as %
\begin{align}
  \hat{\mathbf{E}}^{(+)}(\mathbf{r}_{\rm s},t) &= 
 i\sqrt{\frac{\hbar\beta_{\rm nrad}}{2\epsilon_0\epsilon_I(\mathbf{r}_{\rm s},\omega_c)}}S\tilde{\mathbf{f}}^{\rm N}(\mathbf{r}_{\rm s})\left(a_{\rm out}(t)-a_{\rm in}(t)\right) \nonumber \\
 &\equiv\hat{\mathbf{E}}^{(+)}_{\rm out}(\mathbf{r}_{\rm s},t) -\hat{\mathbf{E}}^{(+)}_{\rm in}(\mathbf{r}_{\rm s},t),  \label{eq:EoutinNRad}
\end{align}
where we have defined the normalized QNM field, $\tilde{\mathbf{f}}^{\rm N}(\mathbf{r}_{\rm s}){=}\sqrt{\omega_c\epsilon_I(\mathbf{r}_{\rm s},\omega_c)/(S\kappa_{\rm nrad)}}\tilde{\mathbf{f}}(\mathbf{r}_{\rm s})$, with respect to integration over the resonator volume $V$, 
where 
\begin{equation}
    \int_{V}{\rm d}\mathbf{r}_{\rm s}|\tilde{\mathbf{f}}^{\rm N}(\mathbf{r}_{\rm s})|^2 = \frac{1}{S^{\rm nrad}}\frac{\omega_c}{\kappa}\int_{V}{\rm d}\mathbf{r}_{\rm s}\epsilon_I(\mathbf{r}_{\rm s},\omega_c)|\tilde{\mathbf{f}}(\mathbf{r}_{\rm s})|^2 = 1,  \label{eq:NormQNMf}
\end{equation}
and in the last step, we used the approximate form of $S^{\rm nrad}$ from Eq.~\eqref{eq:Snradapprox}. We note, that for the source field in the resonator, no further approximations are made, since $\tilde{\mathbf{f}}(\mathbf{r}_{\rm s})$ is a $\omega$-independent function. A schematic diagram of the
important field quantities in the real space is shown
in Fig.~\ref{figschematic}(a), along with the energy levels of the
quantized QNM and TLS, as well as a schematic of the underlying dissipative processes in an abstract quantum picture  
[Fig.~\ref{figschematic}(b)].

%\begin{figure}[th]       
%\centering\includegraphics[clip,trim=0cm 0cm 0cm 0.cm,width=0.99\columnwidth]{SPS_schematic}
 %       \caption{ (a)
  %      Schematic of the system and bath regions for quantization. (b) Energy level digram for the QNM Fock state levels. (b) Energy level diagram for the TLS.\blue{Work in progress - just adding something just now}
 %}
  %      \label{figschematic}
%\end{figure}

Having derived an expression for the output electric field operator, Eq.~\eqref{eq:EoutinRad},~\eqref{eq:EoutinNRad} (in the metal and in the far field), in terms of the source field and an input field, we now define in analogy to the classical power flow,
where we consider the quantum radiative and non-radiative power flow in photon flux units via normal ordered expectation values of the output field:
\begin{align}
p(t) &\equiv \frac{2\epsilon_0}{\hbar\omega_c}\left\{ n_{\rm b}c\int_{S_{\infty}}{\rm d}A_{\mathbf{s}}\langle \hat{\mathbf{E}}^{(-)}_{\rm out}(\mathbf{s},t)\hat{\mathbf{E}}^{(+)}_{\rm out}(\mathbf{s},t)\rangle 
\right . \nonumber \\
 & \left . + \omega_c\int_{V}{\rm d}\mathbf{r}_{\rm s}\epsilon_I(\mathbf{r}_{\rm s},\omega_c)\langle \hat{\mathbf{E}}^{(-)}_{\rm out}(\mathbf{r}_{\rm s},t)\hat{\mathbf{E}}^{(+)}_{\rm out}(\mathbf{r}_{\rm s},t)\rangle\right\},
 \label{eq:pt}
\end{align}
where the first part is the radiative contribution $p^{\rm rad}(t)$ and the second part is the non-radiative contribution $p^{\rm nrad}(t)$. Using the definition of the output field via Eq.~\eqref{eq:EoutinRad}, $p^{\rm rad}(t)$ can be cast into the familiar form 
\begin{align}
    p^{\rm rad}(t)&=S^2 \int_{S_{\infty}}{\rm d}A_{\mathbf{s}}|\tilde{\mathbf{F}}^{\rm N}(\mathbf{R},\omega_{\rm c})|^2 \beta_{\rm rad}\langle a_{\rm out}^{\dagger}(t)a_{\rm out}(t)\rangle  \nonumber \\
    &= S^2 \beta_{\rm rad}\langle a_{\rm out}^{\dagger}(t)a_{\rm out}(t)\rangle,
     \label{eq:ptrad}
\end{align}
where we have  used the normalization condition from Eq.~\eqref{eq:normQNMF}. Since we assume a white noise input in a vacuum state (or/and a classical pump field, which is the self-consistent field pump field that couples directly to the TLA), all normal-ordered expectation values involving input operators $a_{\rm in}$ vanish.
% \magenta{(SF: I think, this is a really important fact and consistent with the Lindblad equations. If the input state is not the vacuum state (for example thermal state), then there are additional dissipators in the Lindblad (such as $\bar{n}(a^{\dagger}\rho a + ...)$ and $(\bar{n}+1)(a\rho a^{\dagger}+...)$), which can also be derived with Ito-Stratonovich calculus, and in addition the normal-ordered expectation values involving $a_{\rm in}$ do not vanish)}. 
Thus,  it follows from Eq.~\eqref{eq:inout} that
\begin{equation}
    p^{\rm rad}(t) = S^2 \kappa\beta_{\rm rad}\langle a^{\dagger}(t)a(t)\rangle=S^2\kappa_{\rm rad}\langle a^{\dagger}(t)a(t)\rangle . 
\end{equation}
Using the normalization condition from Eq.~\eqref{eq:NormQNMf} together with Eq.~\eqref{eq:EoutinNRad} and \eqref{eq:inout}, the same derivation can be applied to $p^{\rm nrad}(t)$, to get
\begin{equation}
     p^{\rm nrad}(t) = S^2 \kappa\beta_{\rm nrad}\langle a^{\dagger}(t)a(t)\rangle=S^2\kappa_{\rm nrad}\langle a^{\dagger}(t)a(t)\rangle .
\end{equation}

To summarize our open-cavity quantization procedure, 
we have adopted a quantized QNM approach
to quantizing modes in metal environments (or indeed arbitrary systems, e.g., with metal and dielectric parts), and extended 
it to derive the spatially-averaged radiative and non-radiative 
decay from first-order quantum correlation functions for photon detection. We have also shown how this connects to standard input-output theory and discussed the symmetrized QNM master equation that must be solved for Fock space quantization.
We stress that our quantization scheme
is not the same as a JC model, as the modes
are quantized with losses; indeed,
 in the multi-mode case, there is
essential non-diagonal coupling between
the QNMs~\cite{franke_quantization_2018}, and the usual multi-mode JC model can completely break down. Moreover, the theory 
shows precisely where the radiative and non-radiative decay processes come from, and also modifies more standard theories
by a factor of $S$ in the power flow. However,
for the example resonator we use below, we find that
$S{\approx} 1$, and so we ignore this prefactor in the evaluation of the correlation functions and photon detection (though it can easily be included or factored into the correlation functions). The extension to compute the second-order quantum correlation function is straightforward, and we simply quote the results below when introducing the relevant figures of merit for
metal-based SPSs. An extension of this input-output theory to handle multiple QNMs will be reported elsewhere.

\section{Figures of merit for Plasmon-Based Single Photon Sources}\label{sec2}

The two essential figures-of-merit of interest for
a SPS include the {\it efficiency} or {\em brightness} of the single photon source and {\it indistinguishability} of the cavity-emitted photons, and we discuss both of these in detail below. 

Using the theory of photon detection \cite{glauber_quantum_1963},
combined with the results above (assuming $S{\approx}1$, which we also compute and justify later), we project on to the
QNM of interest, and define  
% the probability per unit time, per unit volume,
% for  photon emission, is
% \begin{equation}
% p_1 = \braket{E^{(-)}(t,{\bf r}) E^{(+)}(t,{\bf r})},
% \end{equation}
% projected onto the polarization and mode of interest, and at a spatial location 
% of the far field emission.
% Assuming a spatially integrated emission (in the far field), and using
% photon flux units for the output electric field ~\cite{carmichael},
then the probability per unit time
for  photon emission (see Eq.~\eqref{eq:pt}):
\begin{equation}
p_1(t) = \braket{a^\dagger_{\rm out}
a_{\rm out}}(t)\beta_{\rm rad}
= \kappa \braket{a^\dagger
a}(t)\beta_{\rm rad}.
\end{equation}
Thus we can obtain the total emitted photons out of the cavity, as a function of time:
\begin{equation}
P_1^{\rm rad}(t) = \int_{0}^{t}dt'\, \kappa_{\rm rad} \langle a^\dagger a \rangle (t'),
\end{equation}
and for the  photons that 
are non-radiatively absorbed in the metal (Ohmic heating):
\begin{equation}
P_1^{\rm nrad}(t) = \int_{0}^{t}dt'\, \kappa_{\rm nrad} \langle a^\dagger a \rangle (t').
\end{equation}
For convenience, we will also define the total number of photons emitted out, radiatively and non-radiatively:
\begin{equation}
P_1(t) = \int_{0}^{t}dt'\, \kappa \langle a^\dagger a \rangle (t'),
\end{equation}
and so the effect of loss in the metal reduces
the total output emission by the non-radiative 
beta factor, as expected.
Note that  
$\beta_{\rm rad}$ is completely determined from
the properties of the QNM, and both 
$\gamma_{\rm rad}$ and $\gamma_{\rm nrad}$
scale inversely with the QNM effective mode volume.
Note also that
$P_1$ is not necessarily the same as
the photon number probability~\cite{richter_few-photon_2009}, 
$P_1^{\rm num}$ (i.e., the probability for creating a
$n=1$ Fock state), and these are related
by the recursion relationship~\cite{kabuss_inductive_2011} $P_1^{\rm num}{=} P_1 {-} 2!P_2^{\rm num}$, where we have ignored the influence of $P_3^{\rm num} {\approx} 0$. 

In addition to  cavity mode emission,
there is also  spontaneous emission from
background radiation modes, through the
spontaneous emission rate
$\gamma {=} 2d^2 w_a^3 
n_B/(6\pi c^3 \hbar\epsilon_0)$.
 Including the non-radiative decay,
 one  can quantify this difference through the total  $\beta$-factor,
\begin{equation}
\beta = \frac{P_1^{\rm rad}}{P_1+P_a},
\end{equation}
where 
\begin{equation}
P_a(t) = \int_{0}^{t}dt'\, \gamma \langle \sigma^+ \sigma^- \rangle (t'),
\end{equation}
is the mean number of atom-emitted
(or exciton-emitted) photons (i.e., photons emitted into non-cavity modes);
 the temporal
population of the atom and cavity mode is simply $n_a(t) {=} \langle \sigma^+
\sigma^- \rangle(t)$
and $n_c(t) {=} \langle a^\dagger
a\rangle(t)$, respectively.
One thus desires $P_1{\gg} P_a$ and
$\beta_{\rm rad}{>}0.5$, to be dominated by 
radiative SPS in plasmonic resonators.
The values above are implicitly evaluated in
the long time limit, namely
$P_{1/a/2} {=} P_{1/a/2}(t{\rightarrow} \infty)$
(or $P_{1/a/2} {=} P_{1/a/2}(t{\rightarrow} T)$ using the definitions below).

There is also some finite probability of
emitting two photons per pulse, even with a 
$\pi$-pulse excitation, and also in the bad cavity limit.
The joint probability of counting 
two photons,  is~\cite{milburn_quantum_2015,fischer_signatures_2017,Leon2019} 
\begin{align}
P_2 &=  \frac{1}{2}\int_0^{\infty}  dt'' \int_0^{\infty}  dt'
\braket{a^\dagger_{\rm out}(t')
a^\dagger_{\rm out}(t'')
a_{\rm out}(t'')
a_{\rm out}(t')} \nonumber \\
&= 
\frac{1}{2}
\int_0^{\infty}  dt' \int_0^{\infty}  dt''
\kappa^2 \braket{a^\dagger(t') a^\dagger(t'')
a(t'') a(t')} \beta_{\rm rad}^2.
\end{align}
In this case,  total emitted photon
pairs out of the cavity (radiatively), as a function of time,
is
\begin{equation}
P_2^{\rm rad}(t) = \frac{1}{2}\int_{0}^{t} 
dt' \int_{0}^{\infty} dt''
\kappa_{\rm rad}^2 \langle a^\dagger(t') a^\dagger(t''') a(t'')  a(t') \rangle,
\end{equation}
and for non-radiative emission,
\begin{equation}
P_2^{\rm nrad}(t) = \frac{1}{2}\int_{0}^{t} 
dt' \int_{0}^{\infty} dt'' 
\kappa_{\rm nrad}^2 \langle a^\dagger(t') a^\dagger(t'') a(t'')  a(t') \rangle.
\end{equation}
Again, 
for convenience, we will also define the total number of photon pairs emitted, including
into the metal, as
\begin{equation}
P_2(t) = \frac{1}{2}\int_{0}^{t} dt' 
\int_{0}^{\infty}  dt''
 \kappa^2\langle a^\dagger(t') a^\dagger(t'') a(t'')  a (t') \rangle.
\end{equation}
Note that in this case,
$P_2$ as defined is equal
to the  photon number probability,
assuming $P_3^{\rm num} {\approx} 0$ and higher-order photon probabilities also vanish ~\cite{kabuss_inductive_2011,fischer_signatures_2017}.
Importantly, 
we see that the $N$-photon
emissions are reduced by 
$(\beta_{\rm nrad})^{N}$, which helps suppress the ratio of single photons to two photon pairs; however, obviously this could be a serious problem for using metals to generate multi-photon pairs on demand. Note also, that we neglect 
$P_2$ processes via spontaneous emission as they will be negligible in comparison to the cavity (QNM) emitted photon pairs.

% \blue{Comment: Note that Jon Finley's group~\cite{fischer_signatures_2017} have a similar term,  and when they do a $\tau$ integration, they refer to this  as being the probability density for detecting a photon pair with a first emission at time $t$;
% while an integration over $t$ up to $\tau$,
% is termed the probability density for waiting $\tau$ between two emission events.
% This is not so clear to me, but I use the 
% one integrating over $\tau$ for the moment,
% which is why it saturates (in time) so quickly.
% }

In addition to the desire to create one ({\em and only one}) photon out per 
excitation pulse (on-demand), equally as important is to ensure that the emitted photons are
indistinguishable.
The single-photon indistinguishability is a measure of the purity of the quantum state of the emitted photon.
For {\it pulse triggered} photons, the main two detrimental 
effects are from a finite multiphoton probability
(quantified through $G^{(2)}$---defined below---the second-order  correlation functions) as well as pure dephasing, which can spoils the coherence/phase of the emitted photons~\cite{kiraz_quantum-dot_2004}.
One common way to probe the emitter's quantum behavior is through a
Hanbury-Brown-Twiss interferometery setup~\cite{brown1956test}, 
and a 
Hong-Ou-Mandel (HOM) interferometry setup~\cite{hong1987measurement}.
The former can measure the single photon purity, but the latter is important to assess the indistinguishability. 

 For a source where the probability of emitting more than one photon is zero (e.g., a QD which is prepared in the excited state), 
 indistinguishability is a measure of the first-order coherence of the source; the spectrum of each emitted photon is the same as the previous one. For pulse-triggered sources, the multiphoton probability is typically finite, and the quantum state indistinguishability is also a function of the second-order (intensity) coherence. To enable an experimentally-accessible metric of indistinguishability, the phenomena of two-photon interference is usually probed via a HOM interferometry setup. Here, two photons emitted from identical single-photon sources are incident upon a beam splitter, and the cross-correlation function of photodetectors placed at the output channels is measured~\cite{kiraz_quantum-dot_2004,milburn_quantum_2015}.

The HOM experiment is typically
performed with a 
Mach-Zehnder interferometer, 
with a series of pulses with  
 a delay much longer than the lifetime of the exciton~\cite{2fischer16}. To numerically compute the experimentally relevant indistinguishability,
we assume two identical 
QD-cavity systems, whose emitted photons are
combined on a beam splitter~\cite{kiraz_quantum-dot_2004}.
This is equivalent to exciting the  system twice and recombining the photons on a beam splitter as the same time.
The  first and second order coherences
from the two-time correlation functions
(unnormalized), are defined as $G^{(1)}(t,\tau){=}\langle a^\dagger(t)a(t+\tau)\rangle$ and $G^{(2)}(t,\tau){=} \langle a^{\dagger}(t)a^{\dagger}(t+\tau)a(t+\tau)a(t)\rangle$, respectively. The intensity cross-correlation of the output channels $G^{(2)}_{\text{HOM}}(t,\tau)$ is then~\cite{kiraz_quantum-dot_2004,woolley13,2fischer16,gustin_pulsed_2018}
\begin{equation}\label{eq_hom}
G^{(2)}_{\text{HOM}}(t,\tau) = \frac{1}{2}\big(G_{\text{pop}}^{(2)}(t,\tau) + G^{(2)}(t,\tau) - |G^{(1)}(t,\tau)|^2\big),
\end{equation}
where $G_{\text{pop}}^{(2)}(t,\tau){=}\langle a^{\dagger}a\rangle(t)\langle a^{\dagger}a\rangle(t+\tau)$.
We consider a SPS triggered with period $2T$, where $T$ is long enough that the single photon source has returned to its ground state. The indistinguishability 
is then:
\begin{equation}
Ind = 1 - \frac{\int_0^T dt \int_{-T}^T d\tau  G^{(2)}_{\text{HOM}}(t,\tau)}{\int_0^T dt \int_{T}^{3T} d\tau   G^{(2)}_{\text{HOM}}(t,\tau)}.
\end{equation}
When Eq.~\eqref{eq_hom} is time-averaged over $t$, then this is equivalent to taking the ratio of the area on the plot of the cross-correlation of the peak around $\tau {=} 0$ to the peak around $\tau {=} 2T$ and subtracting it from unity~\cite{somaschi16}.

To calculate $Ind$ with only a single pulse excitation (which simplifies the numerical calculation),  for $\tau {>} T$, $G^{(2)}(t,\tau)
{\rightarrow} \langle a^{\dagger}a\rangle(t)\langle a^{\dagger}a\rangle(t+\tau){=} G_{\text{pop}}^{(2)}(t,\tau)$
and $G^{(1)}(t,\tau) {\rightarrow}  \langle a^{\dagger}(t)\rangle
\langle a(t+\tau)\rangle$, and the periodicity of these functions in $t$ can be used to simplify the bounds of integration.
For convenience in separating out the effects
that reduce the indistinguishability, we  next define 
\begin{equation}\label{three}
%\mathcal{I} 
Ind=1 - D_1-D_2,
\end{equation}
where the detrimental effect from
 first order decoherence (but note this function is also affected  by two photon emission) is
 \begin{equation}
%\mathcal{I}
D_1 = \frac{\int_0^T dt \int_{0}^T d\tau  \big(G_{\text{pop}}^{(2)}(t,\tau)
-|G^{(1)}(t,\tau)|^2\big)}{\int_0^T dt \int_{0}^{T} d\tau   \left ( 2G_{\text{pop}}^{(2)}(t,\tau) - |\langle a(t+\tau)\rangle \langle a^{\dagger}(t)\rangle|^2 \right)},
\label{eq:D1}
\end{equation}
 and the detrimental effects from
 second order decoherence (multiphoton emission)
 is 
 \begin{equation}
D_2 = \frac{\int_0^T dt \int_{0}^T d\tau  
 G^{(2)}(t,\tau) \big)}{\int_0^T dt \int_{0}^{T} d\tau   \big(2G_{\text{pop}}^{(2)}(t,\tau) - |\langle a(t+\tau)\rangle \langle a^{\dagger}(t)\rangle|^2\big)}.
 \label{eq:D2}
\end{equation}
  Note that in the limit of a perfectly incoherent single-photon source (e.g., large pure dephasing), that is in the excited state at time $t {=} 0$, this definition of indistinguishability tends to $1/2$, which is similar to experiments.

  For our calculations below, we employ Eqs.~\eqref{three}-\eqref{eq:D2}, evaluating the two-time correlation functions using the quantum regression theorem~\cite{carmichael}.
  For our calculations, we make partial use of the quantum optics toolbox  by Tan~\cite{Tan1999}.
For the time-dependent Rabi field,
we use a Gaussian pulse profile:
\begin{equation}
\Omega(t) {=} \frac{\Theta}{\sqrt{\pi} \tau_{\rm p}} \exp\left [ {-\frac{(t-t_{\rm off})^2}{\tau_{\rm p}^2}}\right ],
\end{equation}
where $\Theta {=} \int_{-\infty}^{\infty} dt\,
\Omega(t)$ is the pulse area, and $\tau_{\text{FWHM}}{=}2\sqrt{\text{ln}(2)}\tau_{\rm p}$.
The offset time $t_{\rm off} {\approx} 3{-}5\, \tau_{\rm p},$ depending on the pulse duration.

\section{Numerical results for gold dimers}\label{sec3}

\subsection{Quasinormal mode parameters
for metal dimer resonators}

\begin{figure}[ht]  
\centering\includegraphics[clip,trim=0cm 0cm 0cm 0.cm,width=0.99\columnwidth]{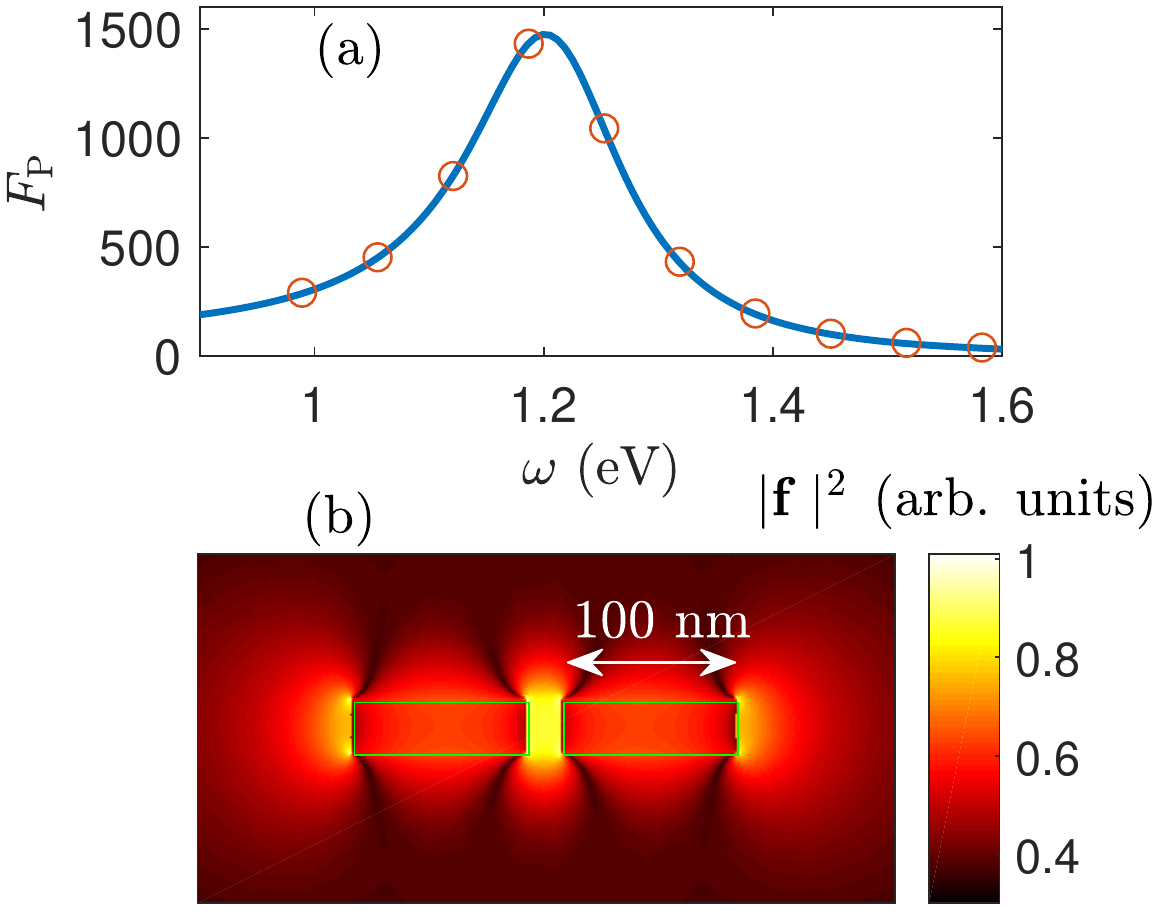}
        \caption{(a) Purcell factor of 
        a point-dipole emitted at the center of a 3D gold nanorod dimer, embedded in a medium with $n_{\rm B}{=}1.5$. The calculations show the single QNM results, which also matches the full dipole simulations~\cite{ge_design_2014,kamandar_dezfouli_regularized_2018} (symbols)
        as well as the quantum theory simulation~\cite{franke_quantization_2018} (which agree within a few \%). The QNM  has a complex QNM frequency 
$\tilde \omega_c {=}1.2067 {-}i 0.0829$~eV. (b) Contour plot of the QNM spatial profile through the center of the dimer structure. The thin green lines show the edge of the gold nanorods, and the gap size is 
        20~nm, with a rod radius of 15~nm.
 }
        \label{figS}
\end{figure}

For the example metal resonator, we consider a 
gold nanorod dimer for the metal resonator,
which is known to produce
good Purcell factors and radiative beta factors~\cite{ge_quasinormal_2014}.
The gold dimer is made up of two nanorods, with a length
$L{=}100~$nm, radius $r{=}15~$nm, and a gap $L_g{=}20~$nm, in a background homogeneous medium with $n_{\rm B}{=}1.5$.
Inside the metal, the dielectric function is described by
the local Drude model,
\begin{equation}
\epsilon_{{\rm metal}}\left(\mathbf{r},\omega\right)=1-\frac{\omega_{p}^{2}}{\omega\left(\omega+i\gamma_{p}\right)},
\end{equation}
where  $\hbar\omega_{p}{=}8.3081\,{\rm eV}$ and $\hbar\gamma_{p}{=}0.0928\,{\rm eV}$ 
for the plasma frequency and collision rate of gold \cite{ge_quasinormal_2014},
respectively.

To compute the QNM,
we adopt the COMSOL technique
in Ref.~\onlinecite{bai_efficient_2013-1}
(see also Ref.~\onlinecite{ge_design_2014} for an FDTD implementation).
In this approach, a search in frequency space is first performed to identify the QNM resonant frequency (pole) and then an additional simulation is performed  near the  resonance frequency to capture the dominant cavity mode of interest. 
The precise numerical details
are described in Ref.~\onlinecite{franke_quantization_2018} (though the dimer design there is slightly different). \
Figure~\ref{figS}(a) shows the QNM Purcell factor
for a dipole position at the dimer center, with a polarization aligned with the QNM (along the gap axis). Figure~\ref{figS}(b) shows the spatial profile of the QNM, and note we can obtain the Purcell factor at any position near or within the dimer structure.

The QNM  has a complex QNM frequency 
$\tilde \omega_c {=}1.2067 {-}i 0.0829$~eV.
Thus the quality factor 
$Q=\omega_c/\kappa{=}7.278$,
%where $\kappa{=}2\gamma_c$,
which offers an enormous bandwidth.
Using a classical dipole simulation,
the quantum efficiency of this 
QNM is $\eta{=} \kappa_{\rm rad}/(\kappa_{\rm rad}{+}\kappa_{\rm nrad}) {=} 0.59  {\approx} 0.6$.
Using the quantum model,
we compute $S_{\rm rad}{\approx} 0.56 \pm 0.025$
and $S_{\rm nrad}{\approx} 0.4 \pm  0.02$, where the estimated uncertainties are from potential numerical uncertainties and errors from the finite gridding and QNM calculations, and spatial integrations~\cite{franke_quantization_2018}; these quantum model calculations yields beta factors in agreement with the
classical model, and shows that
$S{\approx} 1$. The excellent agreement between the classical and quantum QNM results in the bad cavity results for similar such structures are also discussed
in detail in Ref.~\onlinecite{franke_quantization_2018}.
% \blue{Need updated numbers from Mohsen and possibly a graph of classical vs quantum - not sure if this is worth showing though as we can always ref the now similar structure on the PRL arXiv?}
%However, we stress the these classical results are restricted
%to the bad cavity limit.

For the quantum master equation simulations below, 
we  assume a QD dipole moment
of $d{=}30\,$Debye, aligned with the 
QNM polarization, at the center of the dimer.
The corresponding peak Purcell factor is
$F_{\rm{P}}{=}1470$ (cf.~Fig.~\ref{figS}(a)).
However, this value only makes sense in a bad cavity limit and in a Fermi's golden rule, which we cannot use below. Instead, we solve the
QNM system master equation, including the
full pulse dynamics and any time-dependent changes in the 
decay rates.

\subsection{SPS results for a gold nanorod dimer}

Figure~\ref{fig1} below shows the example of the SPS properties with 
$\tau_{\rm p} {=}1/\gamma^{\rm P}$,
which corresponds to 
$\tau_{\rm p}{=}1.76\,$ps (or $\tau_{\rm FWHM} {=} 2.92\,$ps), already shorter than the typical excitation pulses used in SPS experiments.
We assume that the laser is resonant with the QD exciton, on-resonance with the QNM frequency, 
$\omega_{\rm c}$.
Here the decay rate is the Purcell enhanced rate: $\gamma^{\rm P}{=} F_{\rm P} \gamma$.
To assess  the optimum case from radiative dynamics, we first neglect
the role of pure dephasing ($\gamma'{=}0$).
In Fig.~\ref{fig1}(a-b) we show the main two correlation functions responsible for 
degrading the indistinguishability, $D_1$
and $D_2$, and  two-photon emission ($D_2$) is found to be the dominant problem.
We note that we are working in a regime where
$2g/\kappa {=}0.0475$ (and if using a half-width definition
for the cavity decay rate would be $2g/(\kappa/2){=}0.0475$), but we still must solve the quantum dynamics
at the level of a system operator to have two-photon correlations, so the 
QNM operator cannot be adiabatically eliminated (or we would miss many of the effects below with pulse excitation).
Indeed the long-time Purcell factor is not even well defined during the short pulse,
and is reduced during the pulse evolution~\cite{gustin_pulsed_2018}.
For these parameters, even with a short pulse and no pure dephasing, the indistinguishability, $Ind{=}0.773$ (cf.~classical case of 0.5),
and the total brightness
$P_c{=}1.05$, which will then be reduced
further by the quantum efficiency, i.e., by a further 60\%, so $P_c^{\rm rad}{=}1.05 {\times} 0.6 {=} 0.63$. The values of
$2D_1{=}0.162$ and $2D_2{=}0.30$ give the percentage reduction of the indistinguishably to the classical value
of $Ind{=}0.5$; in this simulation regime, the dominant problem is coming from multi-photon emission. Indeed, the only sources of indistinguishability decay considered with these parameters ($\gamma'{=}0$) are multi-photon emission, and decay of $G^{(1)}$ coherence due to emission during the pulse excitation (i.e. see Fig.~\ref{fig1} (a) at small values of $\tau$).
Note for a reduction in pulse width by a factor of two, so
$\tau_{\rm p} {=}0.5/\gamma^{\rm P}$, then we
find $Ind{\approx} 0.85$,
$D_1{=}0.0631$ and $D_2{=}0.0842$, so one requires even shorter pulses
to increase the indistinguishability.

\begin{figure}[th]       
\centering\includegraphics[clip,trim=0cm 0cm 0cm 0.cm,width=0.99\columnwidth]{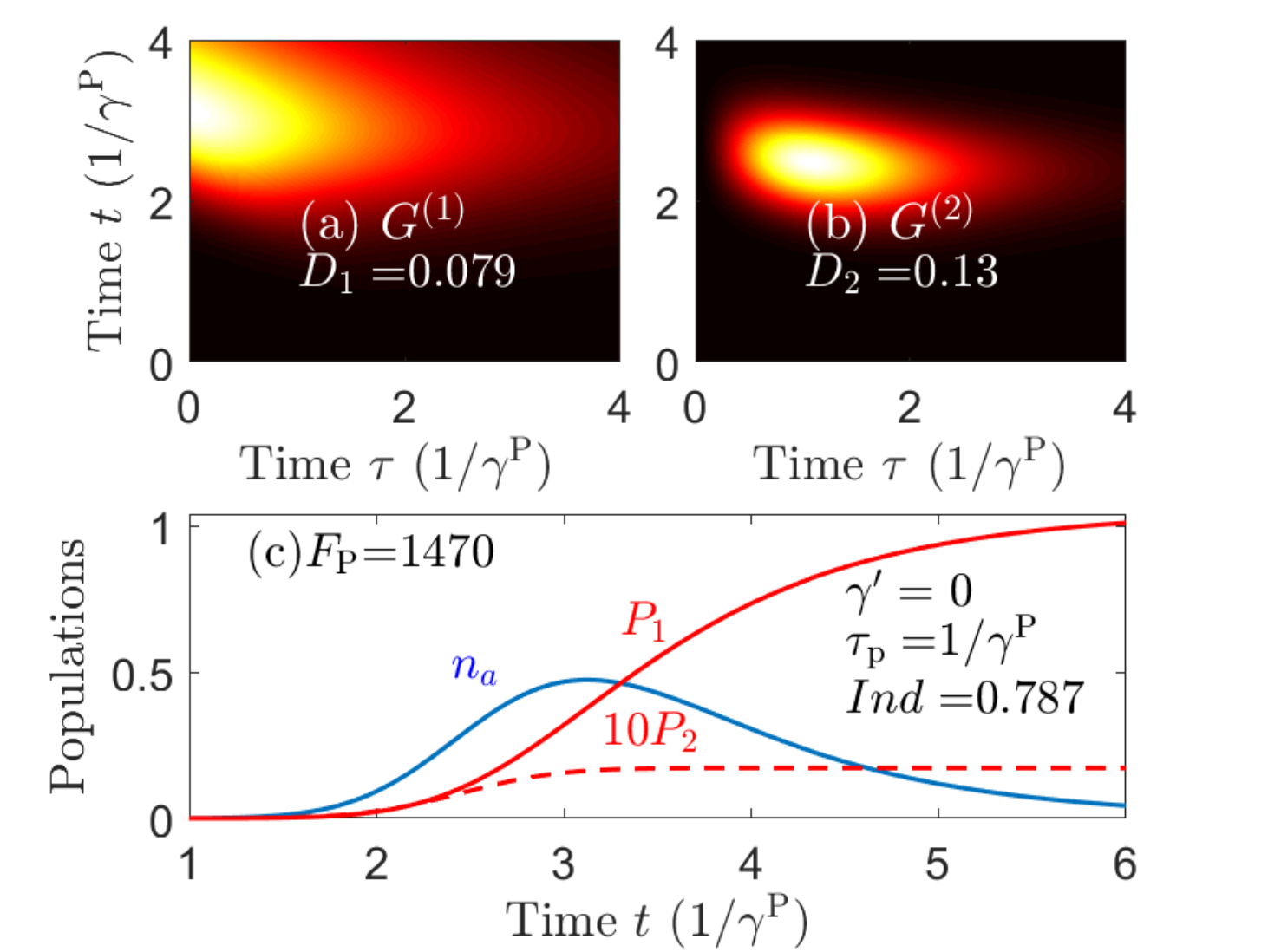}
        \caption{  {SPS simulation  for the gold nanorod, with no pure dephasing, using the Purcell regime  
        $F_{\rm P}{=}1470$ (peak), and $\tau_{\rm p} {=}1/\gamma^{\rm P}$ ($\tau_{\rm FWHM} {=} 2.92\,$ps). The QD dipole moment is 
        30~Debye, which is the same for all cases below, and 
        $2g/\kappa {=}0.0475$.}
 (a) $G^{(1)}(t,\tau)$  function that is integrated for the indistinguishability expression, through Eq.~\eqref{eq:D1}.
 For the colorscale, brighter colors represents the largest values and black represents zero.
 As in (b) but for the $G^{(2)}(t,\tau)$ function that is integrated in the indistinguishability expression, through Eq.~\eqref{eq:D2}. The numbers
 $D_1$ and $D_2$ represent the degradation from first and second order correlations. (c) Populations and efficiency (or brightness) as a function of
 time. We show the total SPS output parameters, and
 note the useful radiation output efficiencies $P_1^{\rm rad}{=}\beta_{\rm rad} P_1$ and $P_2^{\rm rad}{=}\beta_{\rm rad}^2 P_2$, which is related to photons  radiatively emitted to the far field.
}
        \label{fig1}
\end{figure}

\begin{figure}[th]       
\centering\includegraphics[clip,trim=0cm 0cm 0cm 0.cm,width=0.99\columnwidth]{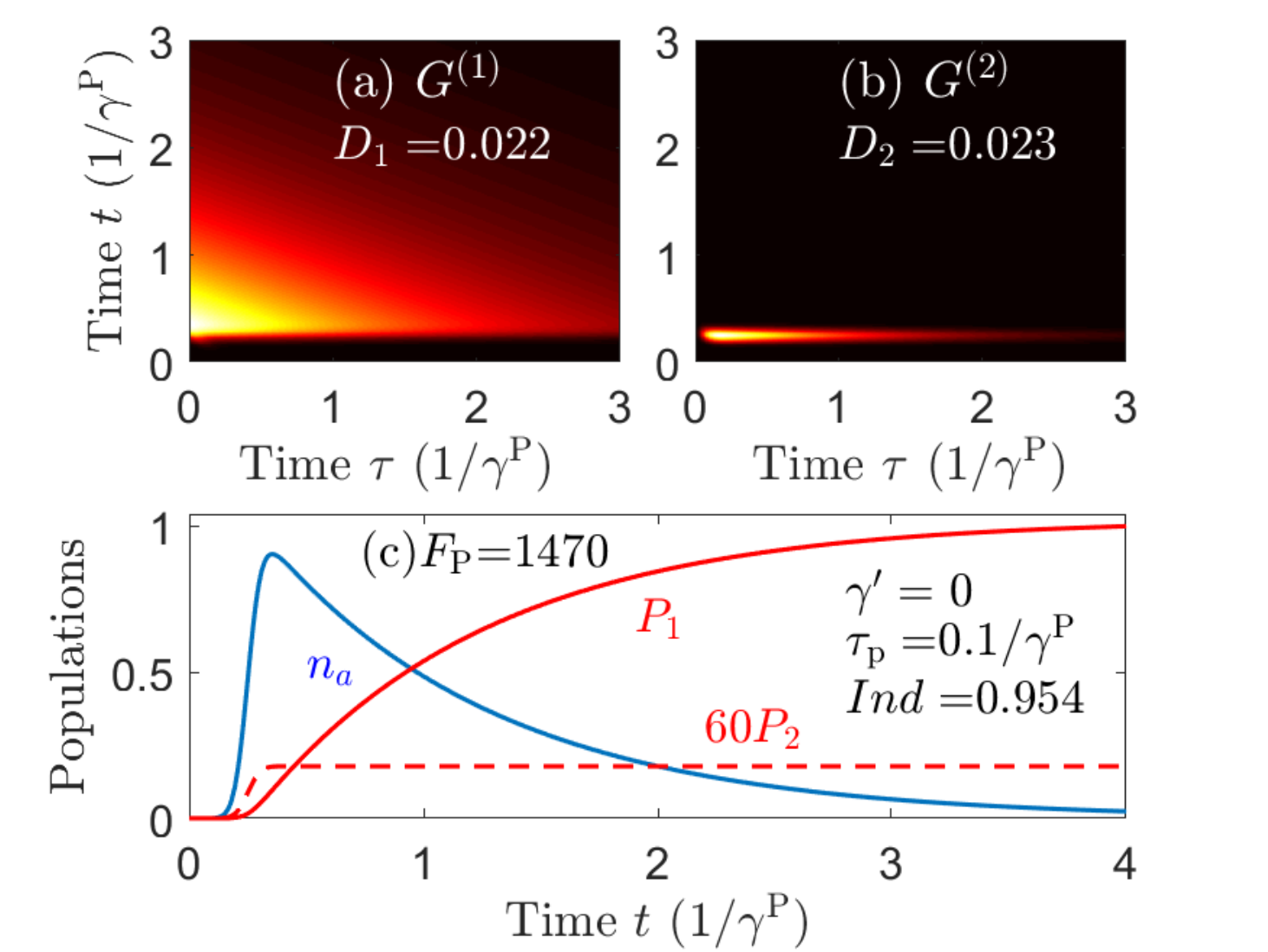}
        \caption{ {SPS simulation  for the gold nanorod, with no pure dephasing, using the  using the Purcell regime  
        {$F_{\rm P}{=}1470$}, and $\tau_{\rm p} {=}0.1/\gamma^{\rm P}$ ($\tau_{\rm FWHM} {=} 0.3\,$ps).}
 (a) $G^{(1)}(t,\tau)$ function that is integrated in the indistinguishability expression, through Eq.~\ref{eq:D1}.
 As in (b) but for the $G^{(1)}(t,\tau)$ function that is integrated in the indistinguishability expression, through Eq.~\eqref{eq:D2}. (c) Populations and efficiency/brightness as a function of
 time. We show the total output parameters, 
 and note that $P_1^{\rm rad}{=}\beta_{\rm rad} P_1$ and $P_2^{\rm rad}{=}\beta_{\rm rad}^2 P_2$.
 }
        \label{fig2}
\end{figure}

Next, in 
Fig.~\ref{fig2},
we reduce the pulse width by a factor of 10, so that
$\tau_{\rm p} {=}0.1/\gamma^{\rm P}$,
and $\tau_{\rm FWHM} {\approx} 0.3\,$ps.
This confirms that sub-ps pulses
are required to yield larger
indistinguishabilities in the high Purcell regime, now yielding
an impressive $Ind{=}0.954$ with
$n_c{=}1.0035$. 
In terms of the total beta factor,
since $P_a {\ll} P_c$ (not shown), the main reduction comes from $\beta_{\rm nrad}$, and
$\beta {\approx} \beta^{\rm rad}$.
We now see that the values of
$2D_1{=}0.044$ and $2D_2{=}0.46$,
have been significantly improved, and in particular the problems of multi-photon emission has been
significantly suppressed by the short pulse.

\begin{figure}[th]       
\centering\includegraphics[clip,trim=0cm 0cm 0cm 0.cm,width=0.99\columnwidth]{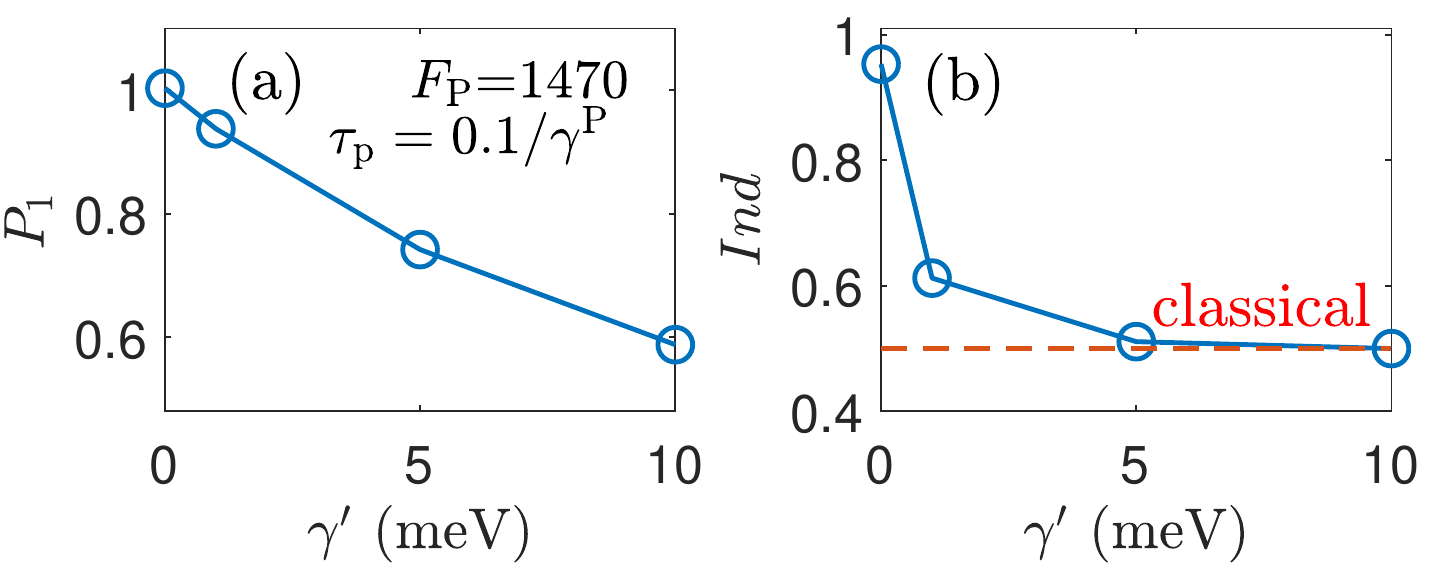}
        \caption{ {SPS figures of merit for the gold nanorod showing the role of TLS pure dephasing,
        using same parameters as in Fig.~\ref{fig2}}.
 (a) Efficiency, $P_1$ (long time limit), versus
 $\gamma'$.
 (b) Indistinguishability, $Ind$, versus  $\gamma'$, showing a significant reduction to the classical value for rates greater than $5\,$meV. We show the total output parameters, and
 the useful radiation output efficiencies $P_1^{\rm rad}{=}\beta_{\rm rad} P_1$.  We show the total output parameters, and  note that $P_1^{\rm rad}{=}\beta_{\rm rad} P_1$ and $P_2^{\rm rad}{=}\beta_{\rm rad}^2 P_2$.
 }
        \label{fig3}
\end{figure}

In reality, the zero phonon line of QDs
and quantum emitters is subject to
 pure dephasing, and thus it is important to asses its role
on the SPS figures of merit. This mechanism can be caused by intrinsic coupling to phonons, and other mechanisms that broaden the zero phonon line (e.g., charge noise~\cite{lodahl_interfacing_2015}). Indeed, many of the claims of plasmon-based SPS generation is the prospect of creating fast SPS at room temperature \cite{hoang_ultrafast_2016-1,livneh_highly_2016}, but, as mentioned earlier, we are not aware of any HOM measurements of these metal-based systems. To assess the role of pure dephasing, in Fig.~\ref{fig3}, we 
use the same parameters as in 
Fig.~\ref{fig2}, but now add in a pure dephasing rate ranging from $1~\mu$eV
to $10\,$meV, corresponding approximately to broadenings  at temperatures of perhaps around 4\,K to 300\,K broadening (though still on the lower side for the room temperature case).
We now see that both the efficiency 
and the indistinguishability are reduced
substantially, and the
indistinguishability actually approaches the classical limit
of 0.5 for the largest pure dephaing rate.
This confirms that room temperature
prospects are likely not feasible, even with these
quite large Purcell factor values, at least not
unless there is a way to reduce the pure dephasing of the zero phonon line at room temperature.

\begin{figure}[th]       
\centering\includegraphics[clip,trim=0cm 0cm 0cm 0.cm,width=0.99\columnwidth]{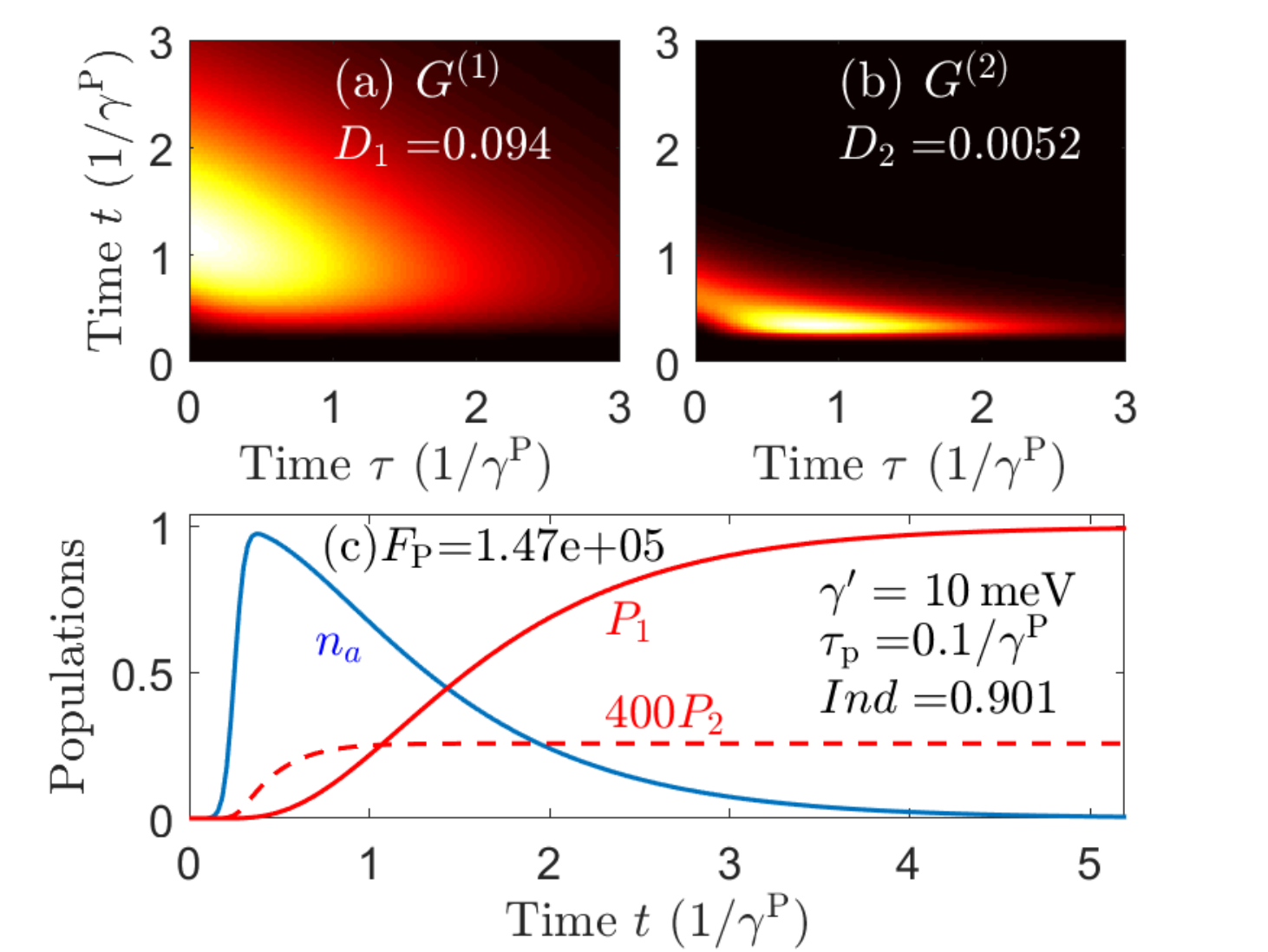}
        \caption{ {SPS simulation with a TLS pure dephasing rate of $\gamma'{=}10\,$meV, but using the much larger Purcell regime
        $F_{\rm P}{=} 1470 {\times}100$, and $\tau_{\rm p} {=}0.1/\gamma^{\rm P}$ ($\tau_{\rm FWHM} {=} 3\,$fs). Note that we now have
        $2g/\kappa {=}0.475$ (i.e., moderate coupling and approaching the good cavity limit).}
 (a) $G^{(1)}(t,\tau)$ function that is integrated in the indistinguishability expression, through Eq.~\ref{eq:D1}.
 As in (b) but for the $G^{(2)}(t,\tau)$ function that is integrated in the indistinguishability expression, through Eq.~\eqref{eq:D2}. (c) Populations and efficiency/brightness as a function of
 time. We show the total output parameters, and  note that $P_1^{\rm rad}{=}\beta_{\rm rad} P_1$ and $P_2^{\rm rad}{=}\beta_{\rm rad}^2 P_2$.
 }
        \label{fig4}
\end{figure}

It is known that the Purcell factor can be increased further, by reducing the gap size.
For example, if we reduce the
gap from 20 nm to 1 nm, then
the Purcell factor exceeds $10^6$ (cf, Ref. \onlinecite{kamandar_dezfouli_quantum_2017});
indeed, one can even reach the vacuum  strong coupling regime at room temperature~\cite{chikkaraddy_single-molecule_2016,van_vlack_spontaneous_2012}, a regime considered to
be not useful for SPS generation (partly due to the expectation of multi-photon states and also the reduction in $\beta^{\rm rad}$). With regards
to the former, however, with ultrashort pulses, the effective coupling rate is reduced during the pulse excitation.  Below we next test the scenario of
a PF that is increased by two orders of magnitude (without doing any specific re-design of the mode parameters), so that
$F_{\rm P}{=}1470{\times}10^{2}$, but we retain the large
pure dephasing value. 
We are now working in a regime with
$2g/\kappa{=}0.475$, and the use of ultrashort pulses where $\tau_{\rm p}\kappa {<}1$ reduces the effective spontaneous emission rate during the pulse below the standard Purcell-enhanced value, increasing the indistinguishability~\cite{gustin_pulsed_2018}. %made edit here -c.g.
Figure~\ref{fig4} summarizes the results using again 
$\tau_{\rm p} {=}0.1/\gamma^{\rm P}$, but note now
that $\tau_{\rm FWHM} {\approx} 3\,$fs. Of course in reality
one will have other energy transitions, but here we want to test the general proof of principle of exciting the TLA with such
an ultrashort pulse, with room temperature pure dephasing levels. Indeed, we see that 
the indistinguishability has been increased again to $Ind{=}0.9$, and the brightness $P_c{\approx} 0.99$. Note if 
$\tau_{\rm p} {=}1/\gamma^{\rm P}$, then
$Ind$ is below 0.6, so the ultrashort pulse is essential to recover good SPS parameters for the indistinguishably. In this latter regime,
more generally, we have shown how one can create 
an efficient SPS in the good cavity limit, a regime that can also be exploited by semiconductor cavities.\\

%\section{Discussion}\label{sec4}
%\red{Might add a general prospects discussion and how to proceed, also with better designs and
%output waveguide channels ...}

% \blue{Comment: It may be worth adding in a discussion (new section - here or earlier) about other
% metal resonators. For example, the metal cube structures
% (e.g.,~\cite{hoang_ultrafast_2016-1})
% are not single mode and rather messy in fact, even though it has been causing quite a lot of experimental interest. We have a detailed QNM analyiss of such a structure and the modes are pretty interesting. The bow tie antennas also have similar Purcell factors (a bit less for the same gap), but larger beta factors; these could be added from the mode and beta viewpoint rather than additional SPS simulations - there is also related fabricated samples with J Finley's group - mentioned in the intro}

\section{Conclusions}\label{sec5}
In summary, we have presented a rigorous quantum optics theory to assess the important parameters for SPSs from
plasmonic resonators. The theory exploits 
a recently developed quantized QNM scheme, where the lossy modes are quantized at the system level, and a single mode master equation is rigorously well defined for simulating multi-photon statistics. The theory was also extended to enable one to compute output fields, which formally separates contributions from radiative and non-radiative reservoirs.
We then introduced a general framework for computing the
key SPS figures of merit, including the efficiency of single photon output per pulse (brightness), as well as the indistinguishability. 

Using an example of a gold nanorod dimer, which is known to yield large Purcell factors and good output beta factors, we first showed how ultrashort pulses are generally required to take advantage of the large Purcell factors (a fact already known for semiconductor cavity systems), which is essential to avoid multi-photon emission during the pulse excitation. Next in the presence of pure dephasing, we showed how the
SPS indistinguishability approaches the classical 
level, when the rates are above $5~$meV, suggesting that room temperature operation is not viable, even though there have been many such claims in the literature. Finally, as a proof of principle, we have shown how to circumvent such a limit, but it requires the use of extremely ultrashort (fs) pulses, where the influence of other exciton states would likely also become problematic. 

It should be stressed that the theory presented is quite general and paves the way for a proper analysis of quantum light sources in dissipative systems such as antenna based single photons and entangled photon emitters, which are too frequently cited as having good figures of merit, but without any proper quantum optical understanding. The need for such a quantized QNM in quantum plasmonics has been highlighted recently~\cite{ACSAsger}.

\vspace{0.5cm}

\section*{Acknowledgements}
We acknowledge Queen's University and the Natural Sciences and Engineering Research Council of Canada for financial support, and CMC Microsystems for the provision of COMSOL Multiphysics to facilitate this research.
We also acknowledge support from the Deutsche Forschungsgemeinschaft (DFG) through SFB 951 Project B12 (Projectnumber 182087777), Project BR1528/8-2 and the
Alexander von Humboldt Foundation through a Humboldt Research Award.

\bibliography{My_Library}
%\bibliography{PRB_old,My_Library,Others}
\end{document}